# ESCAPE

**Preparing Forecasting Systems for the Next generation of Supercomputers**

# D1.3
# Development of Atlas, a flexible data structure framework

Dissemination Level: public

This project has received funding from the European Union's Horizon 2020 research and innovation programme under grant agreement No 67162

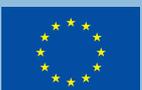

Funded by the
European Union

Co-ordinated by

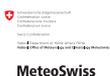 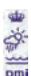 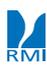 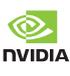 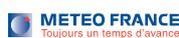 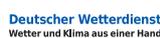 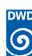 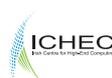 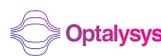 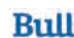 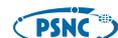 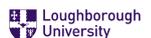

# ESCAPE

**Energy-efficient Scalable Algorithms**
**for Weather Prediction at Exascale**

Author **Willem Deconinck**

Date **31/03/2017**

# Contents

# 1 Executive Summary

The algorithms underlying numerical weather prediction (NWP) and climate models that have been developed in the past few decades face an increasing challenge caused by the paradigm shift imposed by hardware vendors towards more energy-efficient devices. This is because the Dennard scaling (constant power consumption with increasing transistor density) has ended for traditional CPU cores. Rather than increasing clock speeds of the chips, performance is increased by adding more chips, and increasing parallelism. In order to provide a sustainable path to exascale High Performance Computing (HPC), applications become increasingly restricted by energy consumption. As a result, the emerging diverse and complex hardware solutions have a large impact on the programming models traditionally used in NWP software, triggering a rethink of design choices for future massively parallel software frameworks. In this deliverable report, we present *Atlas*, a new software library that is currently being developed at the European Centre for Medium-Range Weather Forecasts (ECMWF), with the scope of handling data structures required for NWP applications in a flexible and massively parallel way. *Atlas* provides a versatile framework for the future development of efficient NWP and climate applications on emerging HPC architectures. The applications range from full Earth system models, to specific tools required for post-processing weather forecast products. The *Atlas* library thus constitutes a step towards affordable exascale high-performance simulations by providing the necessary abstractions that facilitate the application in heterogeneous HPC environments by promoting the co-design of NWP algorithms with the underlying hardware.

*Atlas* provides data structures for building various numerical strategies to solve equations on the sphere or limited area's on the sphere. These data structures may contain a distribution of points (grid) and, possibly, a composition of elements (mesh), required to implement the numerical operations required. *Atlas* can also represent a given field within a specific spatial projection. *Atlas* is capable of mapping fields between different grids as part of pre- and post-processing stages or as part of coupling processes whose respective fields are discretised on different grids or meshes. The latter is particularly relevant for the physical parametrisations, where some physical processes such as radiation may be represented on a coarser grid or mesh and may need to be projected onto a finer grid or mesh.

The key concepts in the design of the *Atlas* data structure are:

- *Grid*: ordered list of points (coordinates) without connectivity rules;

- *Mesh*: collection of elements linking the grid points by specific connectivity rules;

- *Field*: array of discrete values representing a given quantity;

- *FunctionSpace*: discretisation space in which a *field* is defined.

These concepts are depicted in Figure 1, where we used the sphere to represent a global grid, mesh and field. A *grid* is merely a predefined list of two-dimensional points, typically structured and using two indices `i` and `j` so that point coordinates and computational stencils (for e.g. derivatives) are easily retrieved without connectivity rules. For models using a structured grid point approach a *grid* is enough to define *fields* with appropriate indexing mechanisms. For element-based numerical methods (generally unstructured) however, the *mesh* concept is introduced that describes connectivity lists linking elements, edges and nodes.

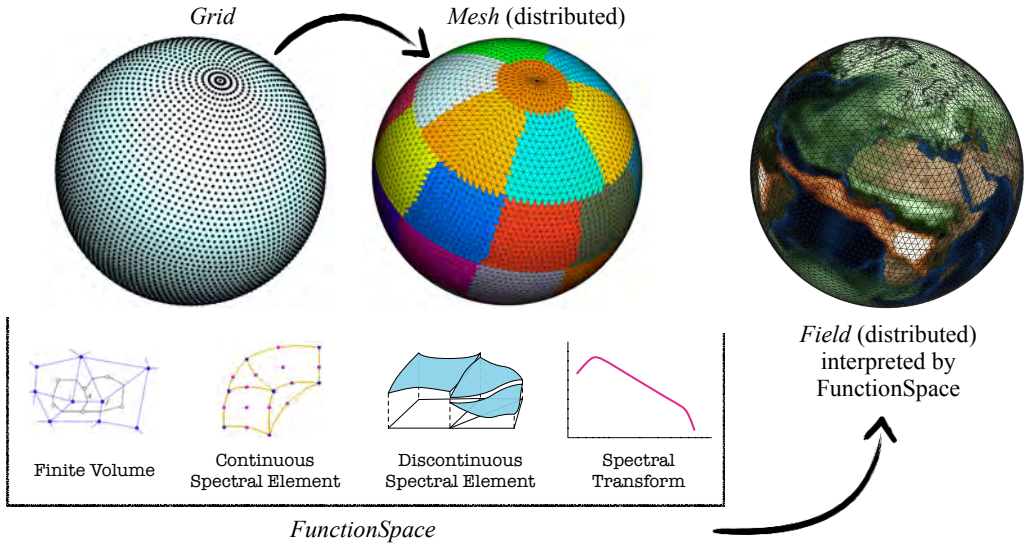

Figure 1: The conceptual design of *Atlas*.

A *mesh* may be decomposed in partitions and distributed among MPI tasks. Every MPI task then allows computations on one such partition. Overlap regions (or halo's) between partitions can be constructed to enable stencil operations in a parallel context.

In addition to these two features, it is necessary to introduce the concept of *field*, intended as a container of values of a given variable. A *field* can be discretised in various ways. The concept responsible to interpret/provide the discretisation of a *field* in terms of spatial projection (e.g. grid-points, mesh-nodes, mesh-cell-centres) or spectral coefficients is the *function space*. The *function space* also implements parallel communication operations responsible for performing synchronisation of fields across overlap regions, which we refer to as halo-exchange hereafter.

A possible *Atlas* workflow consisting of the creation and discretisation of a *field*, is illustrated in Figure 2, where we also emphasise some additional characteristics of each step. The building blocks illustrated in Figure 2 can then be used to implement additional operations required for specific applications. *Atlas* supplies certain mathematical operations as ready solutions to be plugged in to user software. These operations vary from the computation of gradient, divergence, curl and laplacian operations to remapping or interpolation of fields defined on different grids.

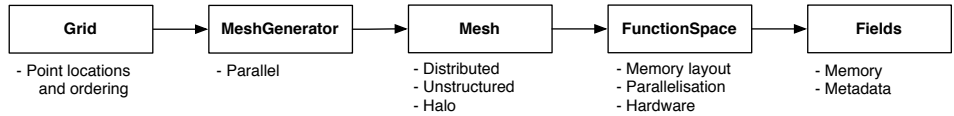

Figure 2: Workflow of *Atlas* starting from *Grid* to the creation of a *Field*, discretised on a *Mesh* and managed by a *FunctionSpace*.

# 2 Introduction

## 2.1 Background

ESCAPE stands for Energy-efficient Scalable Algorithms for Weather Prediction at Exascale. The project will develop world-class, extreme-scale computing capabilities for European operational numerical weather prediction and future climate models. ESCAPE addresses the ETP4HPC Strategic Research Agenda 'Energy and resiliency' priority topic, developing a holistic understanding of energy-efficiency for extreme-scale applications using heterogeneous architectures, accelerators and special compute units by:

- Defining and encapsulating the fundamental algorithmic building blocks underlying weather and climate computing;

- Combining cutting-edge research on algorithm development for use in extreme-scale, high-performance computing applications, minimising time- and cost-to-solution;

- Synthesising the complementary skills of leading weather forecasting consortia, university research, high-performance computing centres, and innovative hardware companies.

ESCAPE is funded by the European Commission's Horizon 2020 funding framework under the Future and Emerging Technologies - High-Performance Computing call for research and innovation actions issued in 2014.

## 2.2 Scope of this deliverable

### 2.2.1 Objectives of this deliverable

The *Atlas* library is a software library being developed at ECMWF in the context of its Scalability Programme. As such, at the initiation of ESCAPE, the library was already in a functional state to support the development of the existing dwarfs (see Deliverable D1.1 [1]). It was however still in an early development stage.

This deliverable aims at providing an established first official and documented release of the *Atlas* library. This release is intended only to be used by ESCAPE partners, with the aim to provide a new stable version of the *Atlas* library to improve ESCAPE *Weather and Climate Dwarfs* (see Deliverable D1.1 [1], and develop new dwarfs (see Deliverable D1.2 [2]).

Most available dwarfs delivered in Deliverable D1.2 embody algorithms defined using domains that span the entire globe. ESCAPE however requires application of these dwarfs to non-global or regional domains. The delivered *Atlas* release therefore also includes new capabilities to accomodate algorithms on regional grids, which have been established in Deliverable D4.4 [3].

Further ESCAPE developments also include the application of a Domain Specific Language (DSL) to several dwarfs. The DSL can have different backends, each capable of executing to execute algorithms on different HPC hardware architectures (CPU, GPU, MIC). Especially GPU architectures are very different in nature and algorithms may require copying data back and forth from a host architecture (CPU) to a device (GPU) where computations on the data are performed (see Deliverable D2.4 [4]). The delivered *Atlas* release therefore also includes a new advanced data-storage facility that accomodates host-device synchronisation capabilities with different backends. In practice the GPU backend is currently implemented only for GPU's programmable with the CUDA language (NVIDIA) [5].

### 2.2.2 Work performed on this deliverable

As suggested in Section 2.2.1, the *Atlas* library was in an early development stage at ESCAPE's initiation. The majority of the work performed during between ESCAPE's initiation and the delivery date has been to design and implement new capabilities as well as redesign and reimplement existing capabilities to accomodate new or evolving requirements. Existing capabilities have been redesigned to make the library easier to use. Other capabilities have been removed and were implemented instead in other more general support libraries *eckit*, *fckit* (see Section 3).

As part of this deliverable, the library *Atlas* has been made to succesfully compile using compiler suites GNU, Intel, Cray and PGI. More specifically, compiling the modern Fortran 2008 interfaces using the PGI compiler suite proved to be not straightforward due to existing compiler bugs. Workarounds in *Atlas* were implemented that allowed PGI's Fortran compiler to compile all of *Atlas* capabilities succesfully. The compiler bugs have been reported to PGI and will be fixed in the upcoming PGI release. Contacts through ESCAPE partner NVIDIA have sped up this process significantly.

ECMWF and ESCAPE partner MeteoSwiss have collaborated to devise a strategy to accomodate the use of *Atlas* as a storage backend for unstructured meshes in the GridTools DSL developed at MeteoSwiss, which will be required for ESCAPE deliverable D2.4 [4]. The GridTools library[6] provides a domain specific language (DSL) that allows to write numerical operators generated from discretisations in a performance portable way, abstracting details of the implementation and optimizations specific to hardware architecture. The ESCAPE deliverable D2.4 will extent the DSL to support unstructured meshes. In order to enable the use of the Atlas unstructured meshes by the DSL, the Atlas data structures have been extended to support GPU accelerators. To further enhance the interoperability between Atlas and the DSL, the GPU support has been implemented using the GridTools storages framework. Additionally the work performed to support Atlas unstructured meshes on GPUs allows to port Fortran dwarfs to GPU using OpenACC.

At the onset of the ESCAPE project, *Atlas* supported mesh generation capabilities for global grids covering the sphere. However as *Atlas* is targeted to be used also in regional NWP models, these capabilites required to be extended for regional grids. This work was mainly done in ESCAPE deliverable D4.4 [3]. Further work on this subject during this deliverable involved consolidating the work performed in deliverable D4.4 and redesigning further features that this major work required.

### 2.2.3   Deviations and counter measures

Even though *Atlas* now fully supports mesh generation for regional grids as required by the majority of limited-area models, there is more work that can be done to support other aspects in *Atlas* such as mathematical operators (gradient, divergence, curl) taking into account the used projections of a regional grid. Although this work is ongoing during the course of the ESCAPE project, it is not foreseen as a critical requirement at this moment to develop algorithms relying on *Atlas* for limited area modelling purposes.

With this deliverable, *Atlas* has been made accelerator aware in terms of data structure. It was envisioned to support also parallel communication operations between accelerators (e.g. GPU's), effectively bypassing the host (CPU). An example would be the support of halo-exchanges between mesh partitions. This support can be seen as an optimisation rather than an obstacle in the development of accelerator aware algorithms relying on *Atlas*. It is therefore not critical for this deliverable.

A new stable *Atlas* release will be delivered for ESCAPE at the end of 2017, which will address these issues. To keep track of the remaining work, it has been added as JIRA tasks in the ESCAPE software collaboration platform [7]. ESCAPE partners will be kept up to date as new features become available in the mean time. This strategy has shown to work effectively over the course of ESCAPE so far.

# 3   Getting started with Atlas

This section is intended to be a general introduction on how to download, install and run *Atlas*. In particular, in section 3.1 we will present the general system requirements before building the library. Section section 3.2 details how to download *Atlas* and its internal dependencies. In section 3.3 we will first describe how to install the internal dependencies required by *Atlas* (if supported by ECMWF) and successively we will outline how to install *Atlas*. Section section 3.4 then explains how to check the installation. Finally, in section 3.5 we show how to incorporate *Atlas* in your own software by creating a simple example that initialises and finalises the library.

## 3.1   System requirements

The system requirements for *Atlas* can be summarised as follows:

- **POSIX**: The operating system must be POSIX compliant. Currently this limits the use to UNIX, Linux, and MacOSX operating systems.

- **C++ 11, Fortran 2008** (optional) : *Atlas* uses the programming languages C++ and optionally Fortran. The required standards for these languages are respectively C++ 11 and Fortran 2008.

- **OpenMP** for C++ (optional): In order for *Atlas* to optionally be able to take advantage of OpenMP multi-threading, the C++ compiler is required to support OpenMP version 3.

- **MPI** for C (optional): To use *Atlas* in a distributed memory application, the system needs to have the MPI libraries for the C-language available.

- **Git**: Required for project management and to download *Atlas*. For use and installation see https://git-scm.com/

- **CMake**: The compilation or build system of *Atlas* is based on CMake 3.3 or higher, which is required to be present on the system. For use and installation see http://www.cmake.org/ .

- **Python**: Required for certain components of the build system. For use and installation see https://www.python.org/. (Known to work with version 2.7.12)

- **Boost** (optional): The *Atlas* installation process can optionally compile unit-tests to check if *Atlas* is correctly installed. To compile these optional unit-tests, the Boost C++ library is required to be present on the system. For use and installation see `http://www.boost.org/`. (Known to work with boost 1.61.0)

- **CUDA** (optional): *Atlas* can also optionally make use of the GridTools storage layer to support use on accelerator hardware. A requirement here is also the Boost C++ library. When intended for a GPU accelerator, an additional requirement is also that CUDA 6.0 or greater be installed on the system.

- **FFTW** (optional): *Atlas* can optionally perform spectral transform operations, which in the most general case require that FFTW be present on the system.

## 3.2   Downloading Atlas

Apart from the system requirements outlined in section 3.1, *Atlas* has a number of internal dependencies that are not all publicly available or require modifications for ESCAPE:

- **ecbuild**: It implements some CMake macros that are useful for configuring and compiling *Atlas* and the other internal dependencies required by *Atlas*. For further information, please visit: `https://software.ecmwf.int/wiki/display/ECBUILD/ecBuild`.

- **eckit**: It implements some useful C++ functionalities widely used in ECMWF C++ projects. For further information, please visit: `https://software.ecmwf.int/wiki/display/ECKIT/ecKit`

- **fckit** (optional): It implements some useful Fortran functionalities.

- **trans, transi** (optional): The *trans* library implements spectral transform methods (in Fortran), and *transi* exposes these methods to be used in C/C++.

- **gridtools_storage** (optional): It implements accelerator-aware data structures.

*Atlas* and the listed internal dependencies are distributed as Git repositories and are available at ECMWF's Bitbucket git hosting service for ESCAPE: `https://software.ecmwf.int/stash/projects/ESCAPE`. The versions for *Atlas* and its internal dependencies are released for this deliverable and tagged in their respective Git repositories with the Git tag "escape/D1.3". Access to this service is currently restricted to ESCAPE partners only. A public access version is to be released with ESCAPE deliverable D2.3 (31 December 2017), including all its dependencies, excluding the optional *trans* and *transi* project.

To download *Atlas* and its internal dependencies, following instructions are to be used on the command line:

```
export ESCAPE=https://software.ecmwf.int/stash/scm/escape
export SRC=$(pwd)/source
mkdir -p ${SRC}
cd ${SRC}
git clone -b escape/D1.3 ${ESCAPE}/ecbuild
git clone -b escape/D1.3 ${ESCAPE}/eckit
git clone -b escape/D1.3 ${ESCAPE}/fckit
git clone -b escape/D1.3 ${ESCAPE}/trans
git clone -b escape/D1.3 ${ESCAPE}/transi
git clone -b escape/D1.3 ${ESCAPE}/gridtools_storage
git clone -b escape/D1.3 ${ESCAPE}/atlas
```

## 3.3  Compilation and Installation of Atlas

In the following we will outline how to build and install *Atlas* and each of the projects *Atlas* depends on that are not covered by the system requirements. The first step is to create a folder where to build and install each project, and to choose a compilation optimisation level. The following three optimisation levels are recommended:

- `DEBUG`: No optimisation - used for debugging or development purposes only. This option may enable additional bounds checking.

- `BIT`: Maximum optimisation while remaining bit-reproducible.

- `RELEASE`: Maximum optimisation. For some algorithms and using some compilers, too agressive optimisation can lead to wrong results.

```
export BUILD=$(pwd)/build
export INSTALL=$(pwd)/install
export BUILD_TYPE=BIT
export PATH=${PATH}:${SRC}/ecbuild/bin
```

```
mkdir -p ${BUILD}/eckit;  cd ${BUILD}/eckit
ecbuild --build=${BUILD_TYPE} --prefix=${INSTALL}/eckit -- ${SRC}/eckit
make -j8 install
```

```
mkdir -p ${BUILD}/fckit;  cd ${BUILD}/fckit
ecbuild --build=${BUILD_TYPE} --prefix=${INSTALL}/fckit -- \
  -DFCKIT_PATH=${INSTALL}/fckit \
  ${SRC}/fckit
make -j8 install
```

```
mkdir -p ${BUILD}/trans;  cd ${BUILD}/trans
ecbuild --build=${BUILD_TYPE} --prefix=${INSTALL}/trans -- \
  ${SRC}/trans
make -j8 install
```

```
mkdir -p ${BUILD}/transi;  cd ${BUILD}/transi
ecbuild --build=${BUILD_TYPE} --prefix=${INSTALL}/transi -- \
  -DENABLE_ESCAPE=ON \
  -DTRANS_PATH=${INSTALL}/trans \
  ${SRC}/fckit
make -j8 install
```

```
mkdir -p ${BUILD}/gridtools_storage;  cd ${BUILD}/gridtools_storage
ecbuild --prefix=${INSTALL}/gridtools_storage -- \
  ${SRC}/gridtools_storage
make -j8 install
```

```
mkdir -p ${BUILD}/atlas;  cd ${BUILD}/atlas
ecbuild --build=${BUILD_TYPE} --prefix=${INSTALL}/atlas -- \
  -DECKIT_PATH=${INSTALL}/eckit \
  -DFCKIT_PATH=${INSTALL}/fckit \
  -DTRANSI_PATH=${INSTALL}/transi \
  -DGRIDTOOLS_STORAGE_PATH=${INSTALL}/gridtools_storage \
  ${SRC}/atlas
```

The following extra flags may be added to Atlas configuration step to fine-tune features

- `-DENABLE_OMP=OFF` — Enable/Disable OpenMP

- `-DENABLE_FORTRAN=OFF` — Disable Compilation of Fortran bindings

- `-DENABLE_TRANS=OFF` — Disable compilation of the spectral transforms functionality. This is automatically disabled if the optional *transi* dependency is not compiled or found. In this case it is also unnecessary to provide `-DTRANSI_PATH=$INSTALL/transi`.

- `-DENABLE_GRIDTOOLS_STORAGE=OFF` — Disable *gridtools_storage*, and enable instead an internal data-storage solution.

- **-DENABLE_GPU=ON** — Enable GPU backend for *gridtools_storage*.

- **-DENABLE_BOUNDSCHECKING=ON** — Enable boundschecking in C++ code when indexing arrays. By default BOUNDSCHECKING is ON when the build-type is DEBUG, otherwise the default is OFF.

**Note**

By default compilation is done using shared libraries. Some systems have linking problems with static libraries that have not been compiled with the flag `-fPIC`. In this case, also compile Atlas using static linking, by adding to the ecbuild step for each project the flag: `--static`

**Note**

The build system for the entire software stack presented above is based on ecbuild which facilitates portability across multiple platforms. However some platforms (like ECMWF's HPC) may have a non-standard configuration (in terms of CMake). For these cases ecbuild has a *toolchain* option, which allows you to provide a custom set of rules for a specific platform. The reader is referred to the ecbuild documentation, and the ecbuild "help" : `ecbuild --help`

The building and installation of *Atlas* should now be complete and you can start using it. With this purpose, in the next section we show a simple example on how to create a simple program to initialise and finalise the library.

## 3.4   Inspecting your *Atlas* installation

Once installation of Atlas is complete, an executable called "atlas" can be found in `${INSTALL}/bin/atlas`. Example use is listed:

```
>>> ${INSTALL}/bin/atlas --version
0.10.0

>>> ${INSTALL}/bin/atlas --git
escape/D1.3

>>> ${INSTALL}/bin/atlas --info
atlas version (0.10.0), git (escape/D1.3)
```

```
Build:
    build type       : Release
    timestamp        : 20160215122606
    op. system       : Darwin-14.5.0 (macosx.64)
    processor        : x86_64
    c compiler       : Clang 7.0.2.7000181
      flags          :  -O3 -DNDEBUG
    c++ compiler     : Clang 7.0.2.7000181
      flags          :  -O3 -DNDEBUG
    fortran compiler : GNU 5.2.0
      flags          :  -fno-openmp -O3 -funroll-all-loops -finline-functions

Features:
    Fortran          : ON
    MPI              : ON
    OpenMP           : OFF
    BoundsChecking   : OFF
    ArrayDataStore   : GridTools
    GPU              : OFF
    Trans            : ON
    Tesselation      : ON
    gidx_t           : 64 bit integer

Dependencies:
    eckit version  (0.12.3), git (escape/D1.3)
    fckit version  (0.3.1), git (escape/D1.3)
    transi version (0.3.2), git (escape/D1.3)
```

This executable gives you information respectively on the version, a more detailed git-version-controlled identifier, and finally a more complete view on all the features that Atlas has been compiled with, as well as compiler and compile flag information. Also printed are the versions of used dependencies such as eckit and transi.

## 3.5   Using *Atlas* in your project

In this section, we provide a simple example on how to link *Atlas* in your own software. We will show a simple "Hello world" program that initialises and finalises the library, and uses the internal *Atlas* logging facilities to print "Hello world!".

Note that *Atlas* supports both C++ and Fortran. Therefore, we will show equivalent examples using both C++ and Fortran.

```cpp
1  // file: hello-world.cc
2
3  #include "atlas/library/Library.h"
4  #include "atlas/runtime/Log.h"
5
6  int main(int argc, char** argv)
7  {
8      atlas::Library::instance().initialise(argc, argv);
9      atlas::Log::info() << "Hello world!" << std::endl;
10     atlas::Library::instance().finalise();
11
12     return 0;
13 }
```

Listing 1: Using *Atlas* in a C++ project

```fortran
1  ! file: hello-world.F90
2
3  program hello_world
4
5  use atlas_module, only : atlas_library, atlas_log
6
7  call atlas_library%initialise()
8  call atlas_log%info("Hello world!")
9  call atlas_library%finalise()
10
11 end program
```

Listing 2: Using *Atlas* in a Fortran project

First, the *Atlas* library is initialised. In C++ this function requires two arguments `argc` and `argv` from the command-line. In Fortran these arguments are automatically provided by the Fortran runtime environment. This function is used to set up the logging facility and for the initialisation of MPI (Message Passage Interface).

Following initialisation, we log "Hello world!" to the `info` channel. *Atlas* provides 4 different log channels which can be configured separately: `debug`, `info`, `warning`, and `error`. By default all log channels print to the std::cout stream, and the debug channel can be switched on or off by setting the environment variable `ATLAS_DEBUG=1` or `ATLAS_DEBUG=0`. Not specifying `ATLAS_DEBUG` is treated as `ATLAS_DEBUG=0`. Finally we end the program after finalising the *Atlas* library.

**Note**

The logging facility exposed by *Atlas* is implemented by *eckit*. The Fortran interface is using *fckit*, which also delegates its implementation to *eckit*. For this reason, logging through C++ or Fortran shares the same infrastructure, which ensures that the logging is consistent in mixed C++/Fortran codes.

**Standard code compilation**

Compiling the C++ example with the GNU C++ compiler:

```
g++ hello-world.cc -o hello-world \
  $(pkg-config ${INSTALL}/atlas/lib/pkgconfig/atlas.pc --libs --cflags)
```

Compiling the Fortran example with the GNU Fortran compiler:

```
gfortran hello-world.F90 -o hello-world \
  $(pkg-config ${INSTALL}/atlas/lib/pkgconfig/atlas.pc --libs --cflags)
```

We can now run the executable:

```
>>> ./hello-world
Hello world!
```

We can run the same executable with debug output printed during Atlas initialisation:

```
>>> ATLAS_DEBUG=1  ./hello-world
```

The output now shows in addition to `Hello world!` also some information such as the version of *Atlas* we are running, the identifier of the commit and the path of the executable, similarly to the output of `atlas --info` in Section 3.4.

**Code compilation using ecbuild**

As *Atlas* is a ecbuild (CMake) project, it integrates easily in other ecbuild (CMake) projects. Two sample ecbuild projects are shown here that compile the "hello-world" example code, for respectively the C++ and the Fortran version.

An example C++ ecbuild project would look like this:

```
1 # File: CMakeLists.txt
2 cmake_minimum_required(VERSION 3.3.2 FATAL_ERROR)
3 project(hello_world CXX)
4
5 include(ecbuild_system NO_POLICY_SCOPE)
6 ecbuild_requires_macro_version(2.6)
7 ecbuild_declare_project()
8 ecbuild_use_package(PROJECT atlas REQUIRED)
9 ecbuild_add_executable(TARGET    hello-world
10                        SOURCES   hello-world.cc
11                        INCLUDES  ${ATLAS_INCLUDE_DIRS}
12                        LIBS      atlas)
13 ecbuild_print_summary()
```

An example Fortran ecbuild project would look like this:

```
1 # File: CMakeLists.txt
2 cmake_minimum_required(VERSION 2.8.4 FATAL_ERROR)
3 project(hello_world Fortran)
4
5 include(ecbuild_system NO_POLICY_SCOPE)
6 ecbuild_requires_macro_version(1.9)
7 ecbuild_declare_project()
8 ecbuild_enable_fortran(MODULE_DIRECTORY ${CMAKE_BINARY_DIR}/module
9                        REQUIRED)
10 ecbuild_use_package(PROJECT atlas REQUIRED)
11 ecbuild_add_executable(TARGET   hello-world
12                        SOURCES  hello-world.F90
13                        INCLUDES ${ATLAS_INCLUDE_DIRS}
14                                 ${CMAKE_CURRENT_BINARY_DIR}
15                        LIBS     atlas_f)
16 ecbuild_print_summary()
```

To compile the ecbuild project, you have to first create an out-of-source build directory, and point ecbuild to the directory where the CMakeLists.txt is located.

```
mkdir -p build; cd build
ecbuild -DATLAS_PATH=${INSTALL}/atlas ../
make
```

Note that in the above command we needed to provide the path to the *Atlas* library installation. Alternatively, **ATLAS_PATH** may be defined as an environment variable. This completes the compilation of our first example that uses *Atlas* and generates an executable into the bin folder (automatically generated by CMake) inside our builds directory. For more information on using ecbuild, or CMake, see https://software.ecmwf.int/wiki/display/ECBUILD/ecBuild.

This completes your first project that uses the *Atlas* library.

# 4   Atlas design and implementation

This section discusses the design of the most important *Atlas* concepts, and to a certain level their implementation details. Implementation details are aided by diagrams formulated in the Unified Modelling Language (UML) [8].

## 4.1   Programming languages

*Atlas* is primarily written in the C++ programming language. The C++ programming language facilitates OO design, and is high performance computing capable. The latter is due to the support C++ brings for hardware specific instructions. In addition, the high compatibility of C++ with C allows *Atlas* to make use of specific programming models such as CUDA to support GPU's, and facilitates the creation of C-Fortran bindings to create generic Fortran interfaces.

With much of the NWP operational software written in Fortran, significant effort in the *Atlas* design has been devoted to having a Fortran OO Application Programming Interface (API) wrapping the C++ concepts as closely as possible.

The Fortran API mirrors the C++ classes with a Fortran derived type, whose only data member is a raw pointer to an instance of the matching C++ class. The Fortran derived type also contains member functions or subroutines that delegate its implementation to matching member functions of the C++ class instance. Since Fortran does not directly interoperate with C++, C interfaces to the C++ class member functions are created first, and it is these interfaces that the Fortran derived type delegates to. The whole interaction procedure is schematically shown in Figure 3. The overhead created by delegating function calls from the Fortran API to a C++ implementation can be disregarded if performed outside of a computational loop. *Atlas* is primarily used to manage the data structure in a OO manner, and the actual field data should be accessed from the data structure before a computational loop starts.

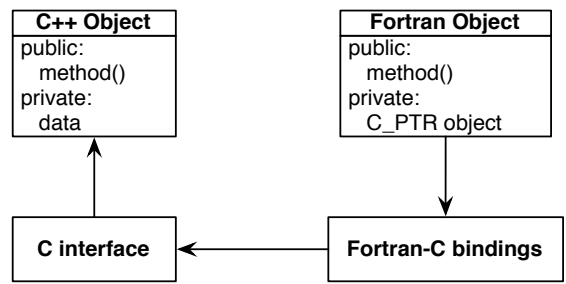

Figure 3: Procedure how the Fortran interface to the C++ design is constructed. When a method in the Fortran object is called, it will actually be executed by the instance of its matching C++ class, through a C interface.

## 4.2 Grid

In the NWP and climate modelling community (as opposed to, for instance, the engineering community) the grid is often a fixed property for a model. One of *Atlas'* goals is to provide a catalogue of a variety of global and regional grids defined by the World Meteorological Organisation in order to support multiple models and model inter-comparison initiatives.

There exist three main categories of grids in terms of functionality that *Atlas* can currently represent: unstructured grids, regular grids, and reduced grids.

Unstructured grids describe an arbitrary number of points in no particular order. The $x$- and $y$-coordinates of the points cannot be computed with certain mathematical formulations, and thus have to be specified individually for each point (e.g. Figure 4a).

Regular grids on the other hand make the assumption that points are aligned in both $x$- and $y$-direction (e.g. Figure 4c). Grid point coordinates can then be derived by two independent indices (`i`,`j`) associated to the $x$- and $y$- direction, respectively.

For reduced grids, lines of constant $y$ or so called parallels may however have a different amount of gridpoints along the $x$-direction (Figure 4b and Figure 4d). Reduced grids are a common type of grid employed in global weather and climate models to reduce the number of points towards the poles in order to achieve a quasi-uniform resolution on the sphere.

For both regular and reduced grids, no assumptions are made on the spacing between the parallels in the $y$ direction. The points in $x$-direction on every parallel are assumed to be equispaced.

*Atlas* provides grid construction facilities based on a configuration object of the type *Config* to create global grids or regional grids. For most global grids, this configuration object can also be inferred from a simple string identifier or *name* containing one or more numbers representing the grid resolution. Commonly used global grids that can currently be accessed through such name are:

- regular longitude-latitude grid (name: `L<NLON>x<NLAT>` or `L<N>`);

- shifted longitude-latitude grid (name: `S<NLON>x<NLAT>` or `S<N>`);

- regular Gaussian grid (name: `F<N>`);

- classic reduced Gaussian grid (name: `N<N>`);

- octahedral reduced Gaussian grid (name: `O<N>`).

In the identifiers shown in this list, `<NLON>` stands for the number of longitudes, `<NLAT>` for the number of latitudes, and `<N>` for the number of parallels between the North Pole and equator (interval `[90°,0°)` ). These grids will be explained in more detail following sections.

Figure 4 showcases 4 example grids that can be created or used with *Atlas*.

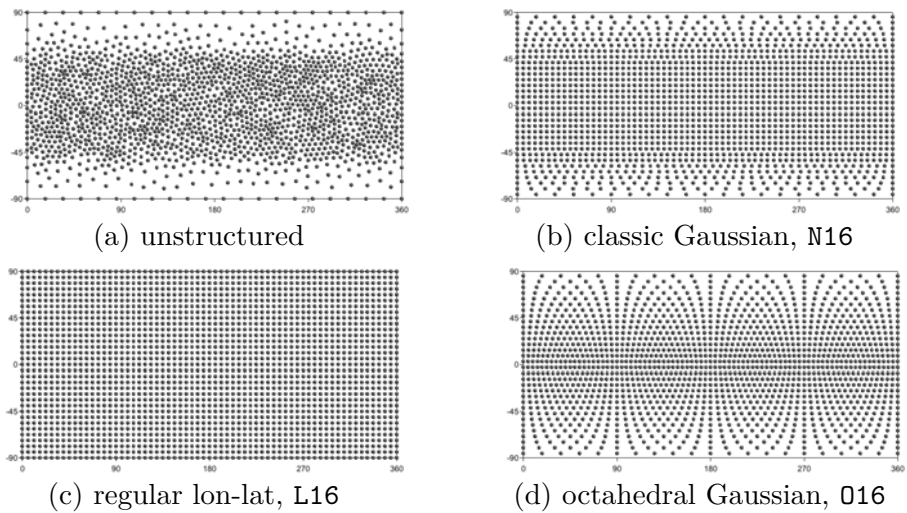

(a) unstructured

(b) classic Gaussian, `N16`

(c) regular lon-lat, `L16`

(d) octahedral Gaussian, `O16`

Figure 4: Four examples of global grids in geographical coordinates with approximately similar resolution in the equatorial region.

### 4.2.1 Projection

In order to support regional grids for the Limited Area Modelling (LAM) community, projections are often needed that transform so called grid coordinates ($x$,$y$) to geographic coordinates (longitude,latitude). For regional grids, the grid coordinates are often defined in meters on a regular grid, as is the case for e.g. a Lambert conformal conic projection and a Mercator projection. Another example projection that is also applicable to a global grid is the Schmidt projection.

In *Atlas*, the projection is embodied by a *Projection* class, illustrated in Figure 5. It wraps an abstract polymorphic *ProjectionImplementation* class with currently 6 concrete implementations:

- LonLat ( type: "lonlat", units: "degrees", identity )

- RotatedLonLat ( type: "rotated_lonlat", units: "degrees" )

- Schmidt ( type: "schmidt", units: "degrees" )

- RotatedSchmidt ( type: "rotated_schmidt", units: "degrees" )

- Mercator ( type: "mercator", units: "meters", regional )

- RotatedMercator ( type: "rotated_mercator", units: "meters", regional )

- Lambert ( type: "lambert", units: "meters", regional )

The *Projection* furthermore exposes functions to convert *xy* coordinates to *lonlat* coordinates and its inverse. For more information about each concrete projection

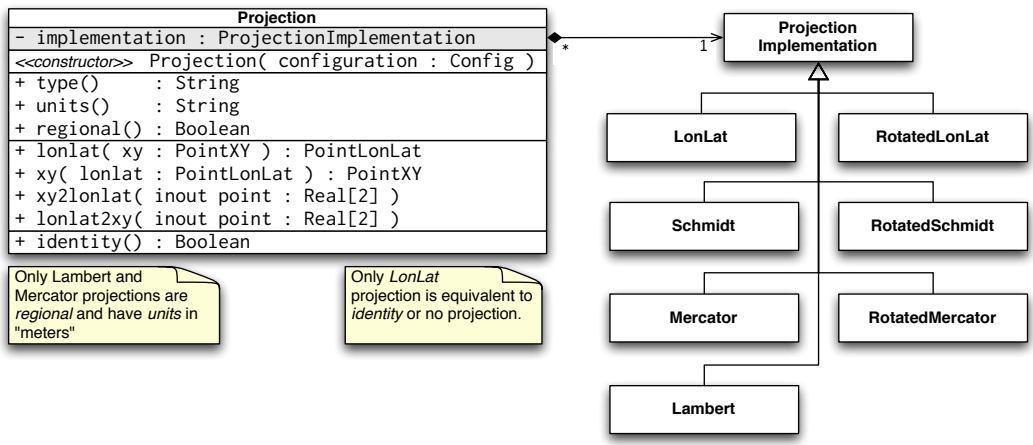

Figure 5: UML class diagram for the *Projection* class

implementation, refer to ESCAPE deliverable report D4.4 [3].

## 4.2.2 Domain

In this section, the *Domain* class is introduced (Figure 6). Its purpose is only useful for non-global grids, and can be used to detect if any coordinate $(x,y)$ is contained within the domain that envelops the grid. The design follows the same principle as the *Projection*: the *Domain* class wraps an abstract polymorphic *DomainImplementation* class with currently 3 concrete implementations:

- Rectangular ( type: "rectangular" )

- ZonalBand ( type: "zonal_band", units: "degrees" )

- Global ( type: "global", units: "degrees" )

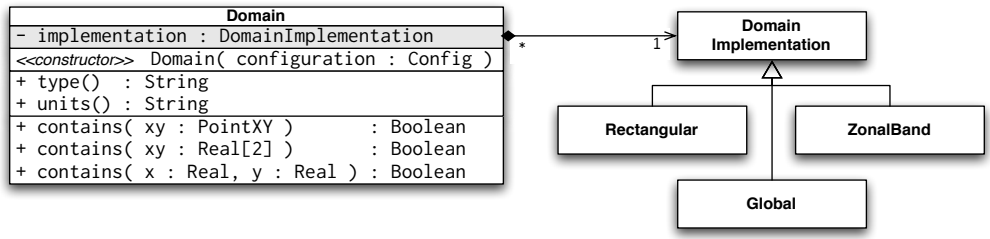

Figure 6: UML class diagram for the *Domain* class

**Note**

The domain has no knowledge of any grid projection. Therefore the points that can be tested to be contained inside the domain must be provided in "grid coordinates" $(x,y)$, and not in geographical coordinates $(lon,lat)$.

The *Rectangular* domain defines a rectangular region defined by 4 values: $x_{\min}$, $x_{\max}$, $y_{\min}$, $y_{\max}$. These values must be defined in units that correspond to the used grid projection. The *ZonalBand* domain assumes that the units of $x$ and $y$ are in degrees, and that the domain is periodic in the $x$-direction. Therefore, to test if a point is contained within this domain only requires to check if the point's $y$ coordinate lies in the interval $[y_{\min}, y_{\max}]$. The *Global* domain, like the *ZonalBand* domain assumes units in degrees, and always evaluates that any point is contained within.

### 4.2.3 Supported Grid types

*Atlas* provides a basic *Grid* class that can embody any unstructured, regular or reduced grid. The *Grid* class is a wrapper to an abstract polymorphic *GridImplementation* class with 2 concrete implementations: *Unstructured* and *Structured*. The *Unstructured* implementation holds a list of $(x,y)$ coordinates (one pair for each grid point). The *Structured* implementation follows the assumption of a reduced grid. It holds a list of $y$-coordinates (one value for each grid parallel), a list of number of points for each parallel, and a list of $x$-intervals (one pair for each parallel) in which the points for the parallel are uniformly distributed. With the *Structured* implementation, both reduced and regular grids can be represented, as regular grids can also be interpreted as a special case of a reduced grid (where every parallel contains the same number of points).

Following code snippets shows how to construct any grid from either a configuration object or a *name*, both in C++ and Fortran.

```cpp
Config F16_config;
F16_config.set( "type", "regular_gaussian" );
F16_config.set( "N", 16 );
Grid F16( grid_config );   // regular Gaussian grid (F16)
Grid N16( "N16" );         // classic reduced Gaussian (N16)
```

Listing 3: Construction of grids, C++ example.

```fortran
type(atlas_Grid)   :: F16, N16
type(atlas_Config) :: F16_config
F16_config = atlas_Config()
call F16_config%set( "type", "regular_gaussian" )
call F16_config%set( "N", 16 )
F16 = atlas_Grid( F16_config )   ! regular Gaussian grid (F16)
N16 = atlas_Grid( "N16" )        ! classic reduced Gaussian grid (N16)
```

Listing 4: Construction of grids, Fortran example

**Note**

Even though the configuration object ( `F16_config` ) is here constructed programatically, it may also be imported through a JSON string or file. The regular Gaussian grid could also be constructed through a name "F16". Similarly the classic reduced Gaussian grid could also be constructed through a config object with the type "classic_gaussian".

Figure 7 illustrates the *Grid* class implementation. It shows that the *Grid* class can return instances of the *Domain* class and the *Projection* class.

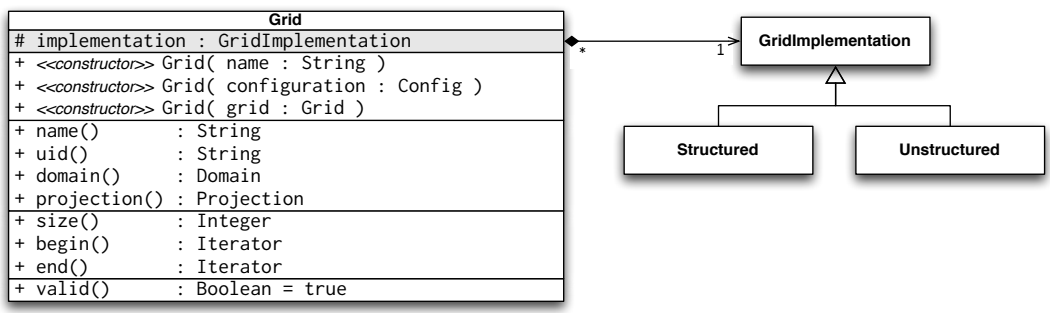

Figure 7: UML class diagram for the *Grid* class

Because this basic *Grid* class can make no assumptions on whether it wraps a *Structured* or a *Unstructured* concrete implementation, it can only expose an interface for the most general type of grids: the *Unstructured* approach. This means that we can find out the number of grid points with the `size()` function, and that we can iterate over all points, assuming no particular order. The following C++ code shows how to iterate over all points, and use the projection to get longitude-latitude coordinates.

```cpp
Grid grid( "O1280" );
Log::info() << "The grid contains " << grid.size() << " points. \n";
for( PointXY p, grid ) {
    Log::info() << "xy: " << p << "\n";
    double x = p.x();
    double y = p.y();

    PointLonLat pll = grid.projection().lonlat(p);
    Log::info() << "lonlat: " << pll << "\n";
    double lon = pll.lon();
    double lat = pll.lat();
}
```

Listing 5: Iterating over all points of a octahedral reduced Gaussian grid (`O1280`)

**Note**

In above C++ code we used the projection to compute the longitude and latitude coordinates. For the used octahedral Gaussian grid however, the projection is of the "lonlat" type by construction, meaning that *x* and *y* are already equivalent to *lon* and *lat* respectively. The second part in the for loop was thus not necessary for this particular grid.

The basic *Grid* class shown in Figure 7 also exposes a function `uid()` which returns a string which is guaranteed to be unique for every possible grid. This includes differences in projections and domains as well.

To be able to expose more structure or properties present in the grid, a number of "grid interpretation" classes are available, that also wrap the used *GridImplementation*, but try to cast it to the *Structured* implementation if necessary. Currently available interpretations classes are:

- *UnstructuredGrid*: The grid is unstructured and cannot be interpreted as structured.

- *StructuredGrid*: The grid may be regular or reduced.

- *RegularGrid*: The grid is regular.

- *ReducedGrid*: The grid is reduced, and *not* regular.

- *GaussianGrid*: The grid may be a global regular or reduced Gaussian grid.

- *RegularGaussianGrid*: The grid is a global regular Gaussian grid.

- *ReducedGaussianGrid*: The grid is a global reduced Gaussian grid, and *not* a regular grid.

- *RegularLonLatGrid*: The grid is a global regular longitude-latitude grid.

- *RegularPeriodicGrid*: The grid is a periodic (in $x$) regular grid.

- *RegularRegionalGrid*: The grid is a regional non-periodic regular grid, and can have any projection.

Note that there is no use case for interpreting a grid as e.g. "octahedral reduced Gaussian" or "classic reduced Gaussian", as it does not bring any benefit over the *ReducedGaussianGrid* interpretation class.

Just like the basic *Grid* class, these interpretation classes have a function `valid()`. Rather than throwing errors or aborting the program if the constraints listed above are not satisfied, the user has to call the `valid()` function to assert the interpretation is possible. Figure 8 illustrates the above list schematically. Arrows indicate a "can be interpreted by" relationship.

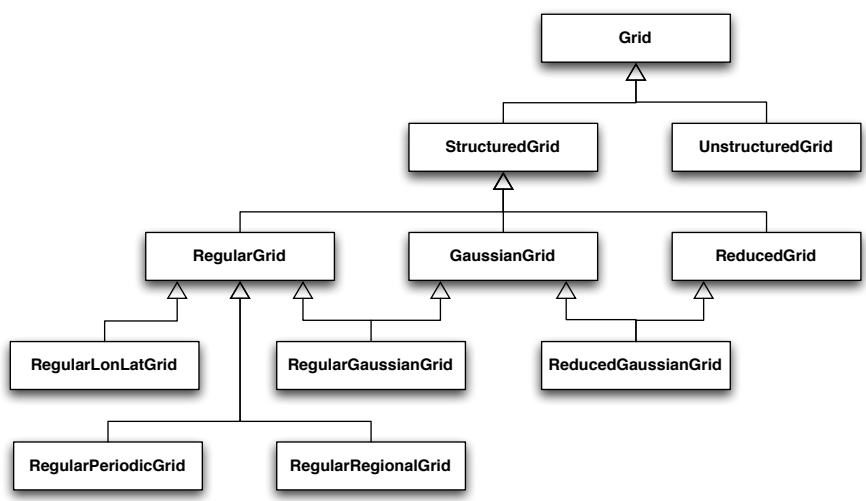

Figure 8: UML class inheritance diagram for *Grid* classes

> **Note**
>
> For a NWP model, you can usually safely assume the grid interpretations as the model can usually only work with a certain type of grid. ECMWF's IFS-model for instance, can assume that all used grids can be interpreted by the *GaussianGrid* class, whereas a LAM-model could e.g. assume the *RegularRegionalGrid* interpretation.

#### 4.2.3.1 UnstructuredGrid

The *UnstructuredGrid* interpretation class constrains the grid implementation to be *Unstructured*. No assumption on any form of structure can be made. Also no assumption on the domain nor the projection used is made.

Figure 9 shows the UML class diagram of the *StructuredGrid*. The first two constructors listed effectively create a new grid, whereas the third constructor accepts any existing grid, and reinterprets it instead. No copy or extra storage is then introduced, since the wrapped *GridImplementation* is a reference counted pointer (a.k.a. `shared_ptr`), of which the reference count is increased and decreased upon *UnstructuredGrid* construction and destruction respectively.

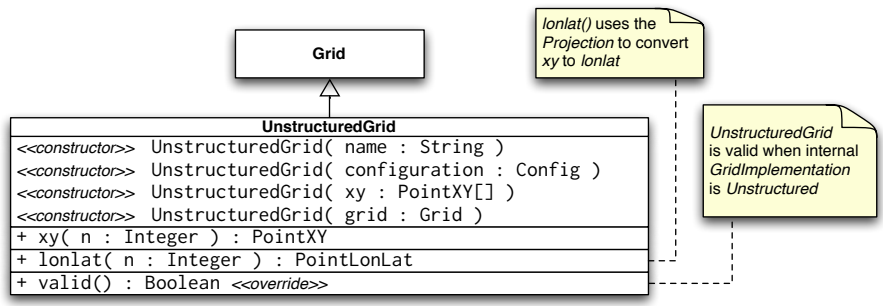

Figure 9: UML class diagram for the *UnstructuredGrid* class

An *UnstructuredGrid* exposes two extra functions `xy(n)` and `lonlat(n)`. The first function gives random access to the $(x,y)$ coordinates of grid point `n`. The second function is a convenience function that internally uses the grid *Projection* to project the grid coordinates `xy(i,j)` to geographic coordinates.

#### 4.2.3.2 StructuredGrid

The *StructuredGrid* interpretation class constrains the grid implementation to be *Structured*. The grid may be regular or reduced. It makes no assumptions on whether the domain is global, periodic, or regional, or whether any projection is used. Almost any grid with some form of structure in a single area can therefore be interpreted by this class.

Figure 10 shows the UML class diagram of the *StructuredGrid*. The first two constructors listed effectively create a new grid, whereas the third constructor accepts any *Grid* , and reinterprets it instead if possible. No copy or extra storage is then introduced, since the wrapped *GridImplementation* is a reference counted pointer (a.k.a. `shared_ptr` ), of which the reference count is increased and decreased upon *StructuredGrid* construction and destruction respectively.

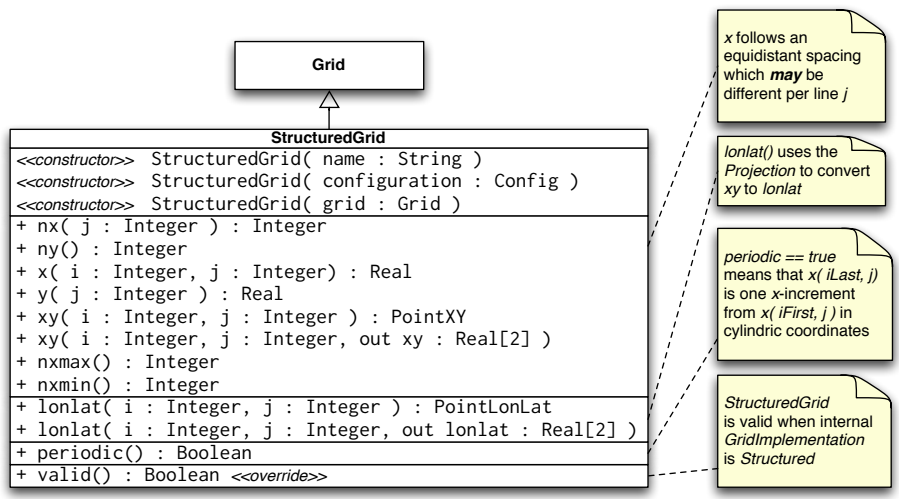

Figure 10: UML class diagram for the *StructuredGrid* class

With the information that the grid can only be reduced or regular, new accessor functions can be exposed to access grid points more effectively through indices (`i`,`j`). The only functions that can be guaranteed to apply for both regular and reduced grids, are the ones that assume a reduced grid. This means that the $x$ coordinate and the number of points on a parallel depend on the parallel itself, denoted by index `j`. For convenience, a function `lonlat(i,j)` is available that internally uses the grid *Projection* to project the grid coordinates `xy(i,j)` to geographic coordinates.

### 4.2.3.3   RegularGrid

A *RegularGrid* is a specialisation of a *StructuredGrid* by further constraining that the number of points on every parallel is equal. In other words, points are now also aligned in $y$ direction. The grid then forms a Cartesian coordinate system.

With this information, access to the $x$ coordinate of a point is now independent of the index `j`, and only depends on the index `i`. The relevant functions that can be adapted now are `nx()` and `x(i)`. Using these functions can possibly increase the performance of algorithms.

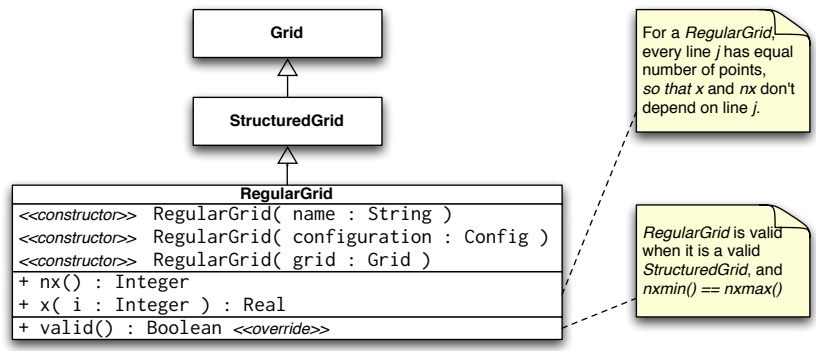

Figure 11: UML class diagram for the *RegularGrid* class

#### 4.2.3.4 ReducedGrid

A *ReducedGrid* is, unlike the *RegularGrid*, not a specialisation of the *StructuredGrid* in terms of functionality, but it does add the constraint that the grid is only valid when it is not regular. Figure 12 shows the class diagram for this type of grid.

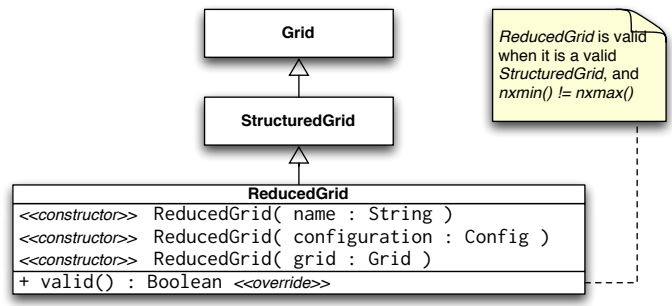

Figure 12: UML class diagram for the *ReducedGrid* class

#### 4.2.3.5 GaussianGrid

A *GaussianGrid* is a *StructuredGrid* with the additional constraint that the grid is globally defined with an even number of parallels that follow the roots of a Legendre polynomial in the interval (90°,-90°) [9]. This class exposes an additional function `N()`, which is the so called Gaussian number, equivalent to the number of parallels between the North Pole and the equator. The $x$-coordinate of each first point of a parallel starts at 0° (Greenwich meridian). Figure 13 shows the class diagram for the *GaussianGrid*.

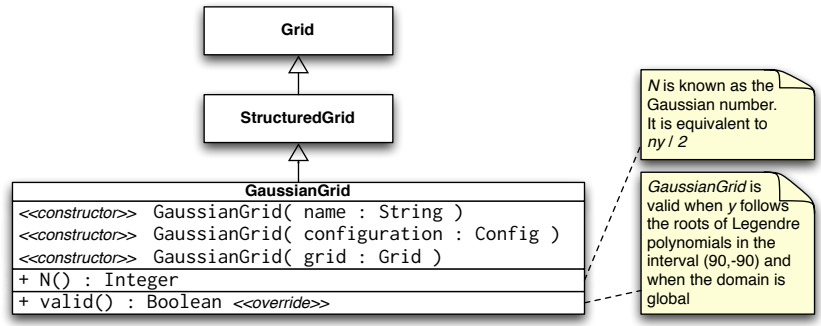

Figure 13: UML class diagram for the *GaussianGrid* class

### 4.2.3.6  RegularGaussianGrid

A *RegularGaussianGrid* combines the properties of a *RegularGrid* and a *Gaussian-Grid*. It can be defined by a single number $N$ (the Gaussian number). The number of points in $x-$ and $y$-direction are by convention

$$nx = 4\ N$$
$$ny = 2\ N$$

Figure 14 shows the class diagram for the *RegularGaussianGrid*. As can be seen in the class diagram, an additional constructor is available, taking only this Gaussian number $N$, so that it is easy to create grids of this type. These grids can also be created through the constructor taking the name "`F<N>`", with `<N>` the Gaussian number $N$.

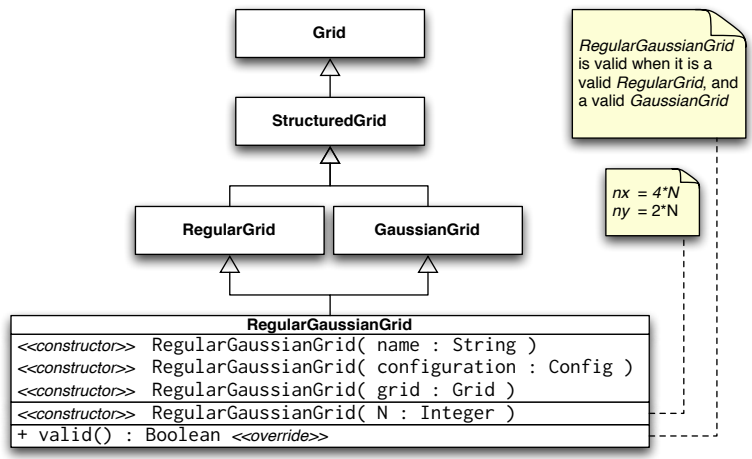

Figure 14: UML class diagram for the *RegularGaussianGrid* class

#### 4.2.3.7 ReducedGaussianGrid

A *ReducedGaussianGrid* combines the properties of a *ReducedGrid* and a *Gaussian-Grid*. A single number $N$ (the Gaussian number), defines the number of parallels ($ny = 2\,N$), but no assumptions are made on the number of points on each parallel.

Figure 15 shows the class diagram for the *ReducedGaussianGrid*. As can be seen in the class diagram, an additional constructor is available, taking an array of integer values with size equal to the number of parallels (must be even). The values correspond to the number of points for each parallel. The WMO GRIB standard also refers to this array as "PL", and IFS refers to this array as "`NLOEN`". In *Atlas* it is referred to as the array $nx$ (cfr. the *StructuredGrid*). The number of parallels $ny$ is inferred by the length of this array, and the Gaussian $N$ number is then $ny/2$, which is used to define the $y$-coordinate of the parallels.

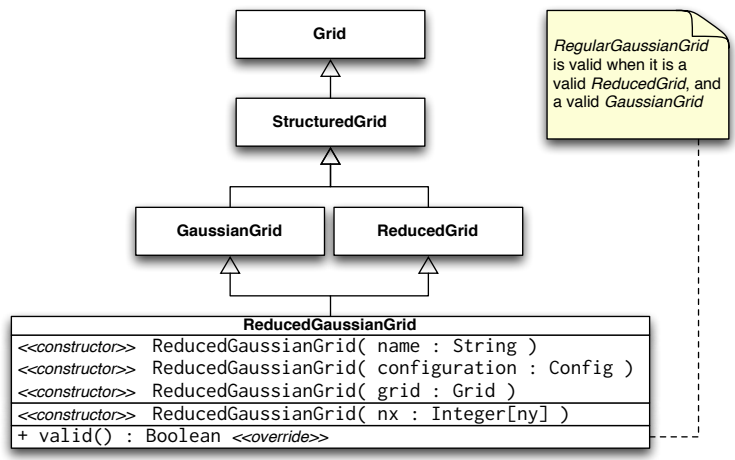

Figure 15: UML class diagram for the *ReducedGaussianGrid* class

#### Classic reduced Gaussian grids

In practise we tend to use only a small subset of the infinite possible combinations of reduced Gaussian grids for a specific $N$ number. Until around 2016, ECMWF's IFS-model was using reduced Gaussian grids for which the $nx$-array was not straightforward to compute. These arrays for all used reduced Gaussian grids were tabulated. We now refer to these grids as "classic" reduced Gaussian grids, and they can be created through the name "N`<N>`", with `<N>` the Gaussian number $N$.

Not any value of $N$ is possible because there are only a limited number of such grids created (only the ones used). *Atlas* can create classic reduced Gaussian grids for values of $N$ in the list [ 16, 24, 32, 48, 64, 80, 96, 128, 160, 200, 256, 320, 400, 512, 576, 640, 800, 1024, 1280, 1600, 2000, 4000, 8000 ].

**Octahedral reduced Gaussian grids**

Since around 2016, ECMWF's IFS-model now uses reduced Gaussian grids for which the $nx$-array can be computed by a simple formula rather than a complex algorithm. These grids are referred to as "octahedral" reduced Gaussian grids. The $nx$-array can be computed as follows in C++:

```cpp
int jLast = 2*N-1;
for( int j=0; j<N; ++j ) {
  nx[j] = 20 + 4*j;      // Up to equator
  nx[jLast-j] = nx[j]; // Symmetry around equator
}
```

Listing 6: Computing the $nx$-array for octahedral reduced Gaussian grids, C++ example

In order to refer to these grids easily in common language, and to more easily construct these grids using the constructor taking a name, the name "O<N>" was chosen, with <N> the Gaussian number $N$, and O referring to "octahedral". The term "octahedral" originates from the inspiration to project a regularly triangulated octahedron to the sphere. Few modifications to the resulting grid were made to make it a suitable reduced Gaussian grid for a spectral transform model [10].

**Note**

Models or other software applications should not treat the octahedral reduced Gaussian grid as a special case. For all means and purposes it is still a reduced Gaussian grid, following all requirements layed out by the WMO GRIB standard!

#### 4.2.3.8 RegularLonLatGrid

The *RegularLonLatGrid* is likely the most commonly used grid on the sphere. It is a global grid regular grid defined in degrees with a uniform distribution both in $x$- and in $y$-direction. *Atlas* supports 4 variants of the *RegularLonLatGrid*, each with 2 identifier names:

- standard: `L<NLON>x<NLAT>` or `L<N>`

- shifted: `S<NLON>x<NLAT>` or `S<N>`

- longitude-shifted: `SLON<NLON>x<NLAT>` or `SLON<N>`

- latitude-shifted: `SLAT<NLON>x<NLAT>` or `SLAT<N>`

In the identifier names, `<NLON>` and `<NLAT>` denote respectively $nx$ and $ny$ of a regular grid. For ease of comparison with the Gaussian grids, these grids can also be named instead with a $N$ number denoting the number of parallels in the interval $[90°, 0°)$ – between the North Pole and equator by including Pole and excluding equator. The $x$- and $y$-increment is then computed as $90°/N$. For each of the grids, all points are defined in the range $0° \leq x < 360°$ and $-90° \leq y \leq +90°$. For the *standard* case, the first and last parallel are located exactly at respectively the North and South Pole. Usually the number of parallels $ny =$ `<NLAT>` is odd, so that there is also exactly one parallel on the equator. It is also guaranteed that the first point on each parallel is located on the Greenwich meridian ($x = 0°$). In this context, *shifted* denotes a shift or displacement of $x$- and $y$-coordinates of all points with half increments with respect to the standard (or unshifted) case. In order to achieve the same $x$- and $y$-increment as the *standard* case, the *shifted* case should be constructed with one less parallel. The two remaining cases *longitude-shifted* and *latitude-shifted* shift only respectively the $x$ or $y$ coordinate of each grid point.

Figure 16 shows the class diagram for the *RegularLonLatGrid*. It can be seen that this class exposes 4 functions to query which of the 4 variants is presented.

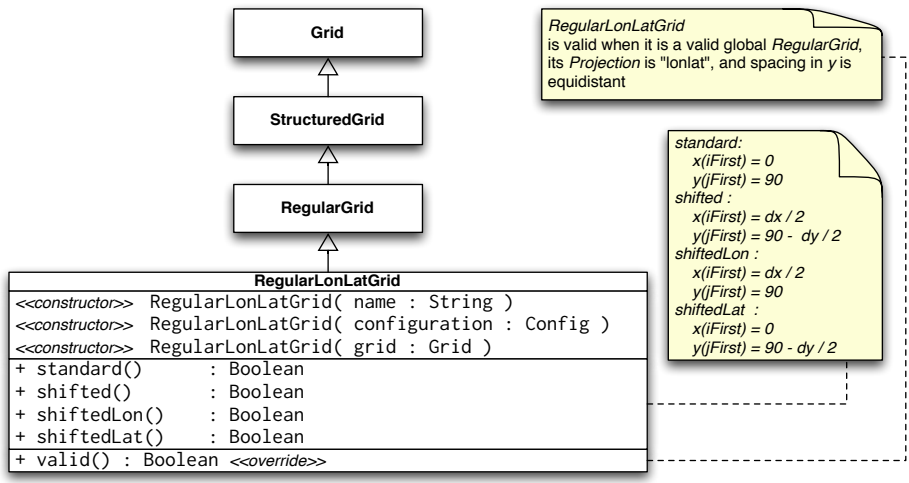

Figure 16: UML class diagram for the *RegularLonLatGrid* class

### 4.2.3.9 RegularPeriodicGrid

The *RegularPeriodicGrid* can be used to assert that the grid is a regular grid with equidistant spacing in $x$- and $y$-direction, and with periodicity in the $x$-direction. The latter enforces an implicit additional constraint that $x$ and $y$ are defined in degrees. Figure 17 shows the class diagram for the *RegularPeriodicGrid*.

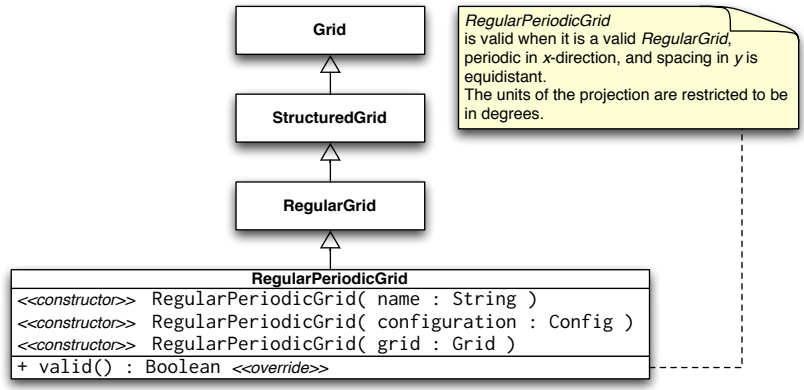

Figure 17: UML class diagram for the *RegularPeriodicGrid* class

### 4.2.3.10 RegularRegionalGrid

The *RegularRegionalGrid* is a grid that asserts that the grid is not global nor periodic. The gridpoints must be equidistant both in $x$- and $y$-direction. No restrictions on projections are made. This grid would be the typical use-case grid to use in conjuction with e.g. a Lambert, Mercator, or RotatedLonLat projection. Figure 18 shows the class diagram for the *RegularRegionalGrid*. Construction of grids of this type can be done in various ways through configuration. Refer to ESCAPE deliverable report D4.4 [3] for more information.

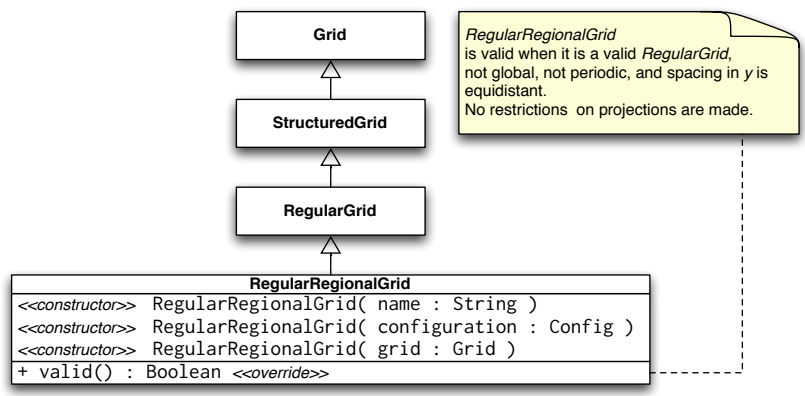

Figure 18: UML class diagram for the *RegularRegionalGrid* class

### 4.2.4 Partitioner

Even though the *Grid* object itself is not distributed in memory as it does not have a large memory footprint, it is necessary for parallel algorithms to divide work over parallel MPI tasks.

There exist various strategies in how to partition a grid, where each strategy may offer different advantages, depending on the grid and numerical algorithms to be used.

*Atlas* implements a grid *Partitioner* class, that given a grid, partitions the grid and creates a *Distribution* object that describes for each grid point which partition it belongs to. Figure 19 illustrates the UML class diagram for the *Partitioner* class. Following a similar design philosophy as before, the *Partitioner* class wraps an abstract polymorphic *PartitionerImplementation* object. Figure 20 illustrates the UML class diagram for the *Distribution* class.

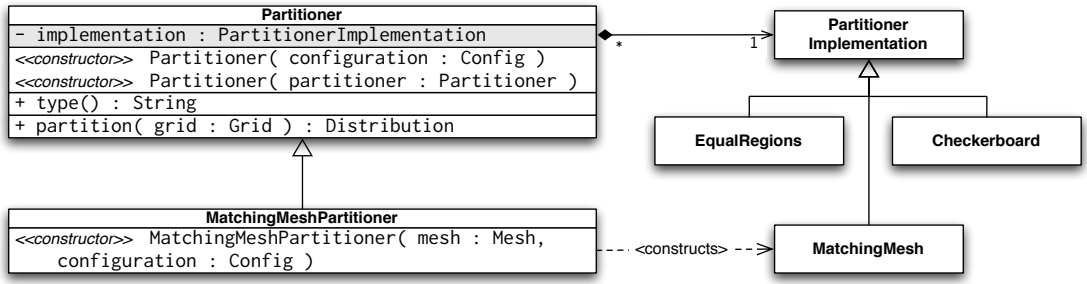

Figure 19: UML class diagram for the *Partitioner* class

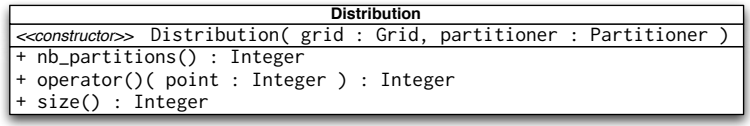

Figure 20: UML class diagram for the *Distribution* class

Currently there are 3 concrete implementations of the *PartitionerImplementation*:

- Checkerboard ( type: "checkerboard" ) – Partitions a grid in regular zones

- EqualRegions ( type: "equal_regions" ) – Partitions a grid in equal regions, reminiscent of a disco ball.

- MatchingMesh ( type: "matching_mesh" ) – Partitions a grid such that grid points following the domain decomposition of an existing mesh which may be based on a different grid.

The *Checkerboard* and *EqualRegions* implementations can be created from a configuration object only. The *MatchingMesh* implementation requires a further mesh argument to its constructor. For this reason, a *MatchingMeshPartitioner* class exists whose only purpose is that it knows how to construct its related *MatchingMesh* implementation with the extra mesh argument.

### 4.2.4.1 Checkerboard Partitioner

For regular grids, such as the one depicted in Figure 4c, a logical domain decomposition would be a checkerboard. The grid is then divided as well as possible into approximate rectangular zones in Cartesian grid coordinates $(x,y)$ with an equal number of grid points. An example of this partitioning algorithm is shown in Figure 21.

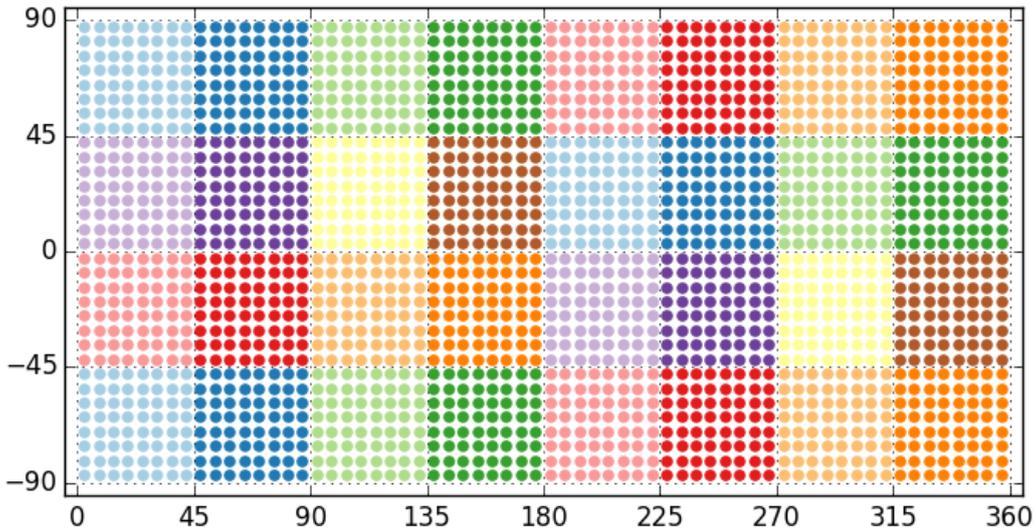

Figure 21: Example *Checkerboard* partitioning of a shifted regular longitude-latitude grid (`S64x32`) in 32 partitions.

### 4.2.4.2 EqualRegions Partitioner

For reduced grids as the ones shown in Figure 4b and Figure 4d or for uniformly distributed unstructured grids, an "equal regions" domain decomposition is more advantageous [11]–[13]. The "equal regions" partitioning algorithm divides a two-dimensional grid of the sphere (i.e. representing a planet) into bands from the North pole to the South pole. These bands are oriented in zonal directions and each band is then split further into regions containing equal number of grid points.

The only exceptions are the bands containing the North or South Pole, that are not subdivided into regions but constitute North and South polar caps.

An example of this partitioning algorithm is shown in Figure 22

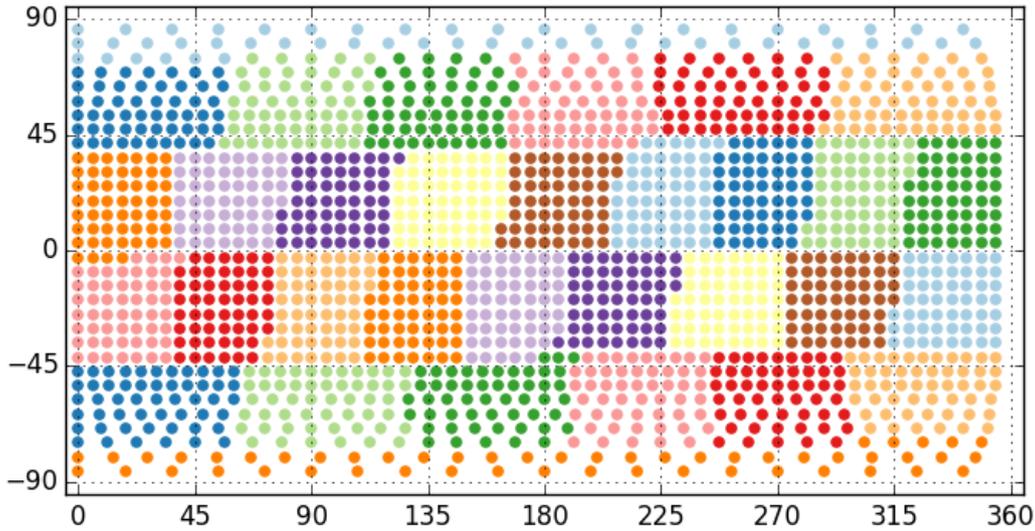

Figure 22: Example *EqualRegions* partitioning of a `N16` classic reduced Gaussian grid in 32 partitions.

#### 4.2.4.3    MatchingMesh Partitioner

The *MatchingMeshPartitioner* allows to create a *Distribution* for a grid such that the grid points follows the domain decomposition of an existing mesh (described in detail in Section 4.3). This partitioning strategy is particularly useful when grid points of a partition should be contained within a mesh partition present on the same MPI task to avoid parallel communication during coupling or interpolation algorithms. Note that there is no guarantee of any load-balance here for the partitioned grid. Figure 23 shows an example application of the *MatchingMeshPartitioner*.

### 4.3    Mesh

For a wide variety of numerical algorithms, a *Grid* (i.e. a mere ordering of points and their location) is not sufficient and a *Mesh* might be required. This is usually obtained by connecting grid points using polygonal elements (also referred to as cells), such as triangles or quadrilaterals. A mesh, denoted by $\mathcal{M}$, can then be

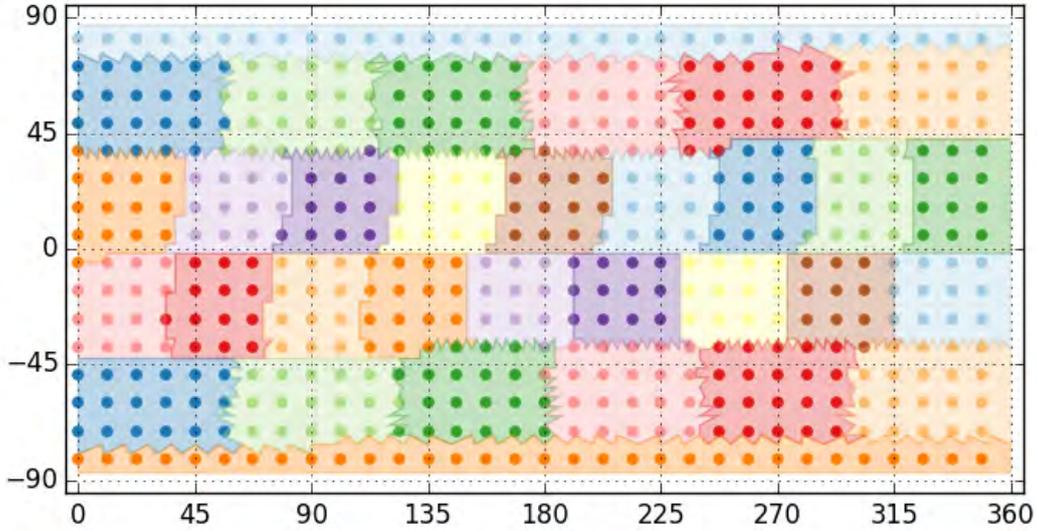

Figure 23: Example partitioning in 32 parts of a F8 rectangular Gaussian grid (solid dots) using the domain decomposition of an existing meshed N24 classic reduced Gaussian grid. Each domain is shaded and surrounded by a solid line. The jagged lines of the existing N24 mesh subdomains are contours of its elements.

defined as a collection of such elements $\Omega_i$:

$$\mathcal{M} := \cup_{i=1}^{N} \Omega_i .$$ (1)

For regular grids, the mesh elements can be inferred, as a blocked arrangement of quadrilaterals. For unstructured grids or reduced grids (Section 4.2), these elements can no longer be inferred, and explicit connectivity rules are required. The *Mesh* class combines the knowledge of classes *Nodes*, *Cells*, *Edges*, and provides a means to access connectivities or adjacency relations between these classes). *Nodes* describes the nodes of the mesh, *Cells* describes the elements such as triangles and quadrilaterals, and *Edges* describes the lines connecting the nodes of the mesh. Figure 24 sketches the composition of the *Mesh* class with common access methods for its components. Differently from the *Grid*, the *Mesh* may be distributed in memory. The physical domain $S$ is decomposed in sub-domains $S_p$ and a corresponding mesh partition $\mathcal{M}_\mathtt{p}$ is defined as:

$$\mathcal{M}_\mathtt{p} := \{\cup\, \Omega\ ,\quad \forall\ \Omega\ \in\ \mathcal{S}_\mathtt{p}\}. \tag{2}$$

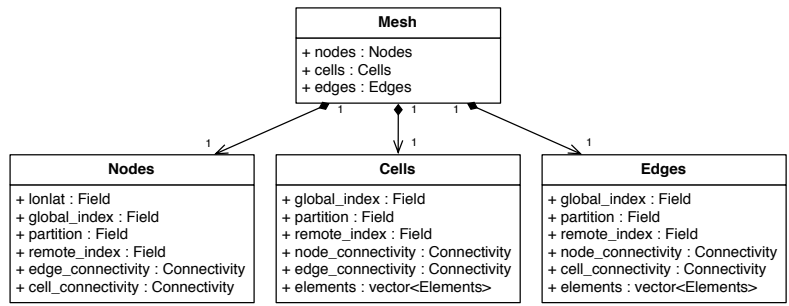

Figure 24: Mesh composition

More details regarding this aspect are given in Section 4.4.

A *Mesh* may simply be read from file by a *MeshReader*, or generated from *Grid* by a *MeshGenerator*. The latter option is illustrated in Figure 2, where the grid points will become the nodes of the mesh elements. Listing 7 shows how this can be achieved in practice, and Figure 25b visualises the resulting mesh for grids `N16` and `O16`.

```
Grid          grid( "O16" );
MeshGenerator generator( "structured" );
Mesh          mesh = generator.generate( grid );
```

Listing 7: C++ Mesh generation from a *StructuredGrid*

**Note**

For *UnstructuredGrid*s, another *Meshgenerator* needs to be used based on e.g. Delaunay triangulation (type="delaunay"). Whereas the *StructuredMeshGenerator* is able to generate a parallel distributed mesh in one step, the *DelaunayMeshGenerator* currently only supports generating a non-distributed mesh using one MPI task. In the future it is envisioned that this implementation will be parallel enabled as well.

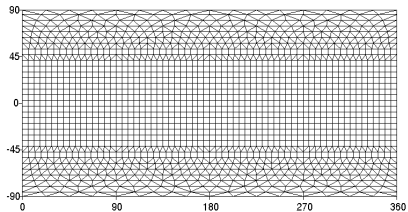

(a) classic Gaussian, `N16`

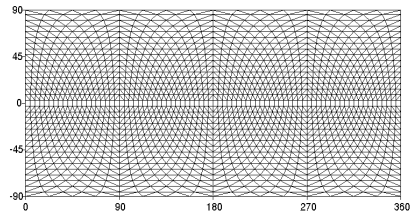

(b) octahedral Gaussian, `O16`

Figure 25: *Mesh* generated for two types of reduced grids (Figure 4)

Because several element types can coexist as cells, the class *Cells* is composing a more complex interplay of classes, such as *Elements*, *ElementType*, *BlockConnectivity*, and *MultiBlockConnectivity*. This composition is detailed in Figure 26. Atlas provide various type of connectivity tables: BlockConnectivity, IrregularConnectivity and MultiBlockConnectivity. BlockConnectivity is used when all elements of the mesh are of the same type, while IrregularConnectivity is more flexible and used when the elements in the mesh can be of any type. The BlockConnectivity implementation has a regular structure of the lookup tables and therefore provides better computational performance compared to the IrregularConnectivity. Finally the MultiBlockConnectivity supports those cases where the mesh contains various types of elements but they can still be grouped into collections of elements of the same type so that numerical algorithms can still benefit from performing operations using elements of one element type at a time. The *Elements* class provides the view of elements of one type with node and edge connectivities as a *BlockConnectivity*. The interpretation of the elements of this one type is delegated to the *ElementType* class. The *Cells* class is composed of multiple *Element*s and provides a unified view of all elements regardless of their shape. The *MultiBlockConnectivity* provides a matching unified connectivity table. Each block in the MultiBlockConnectivity shares its memory with the BlockConnectivity present in the *Element*s to avoid memory duplication (see Figure 27).

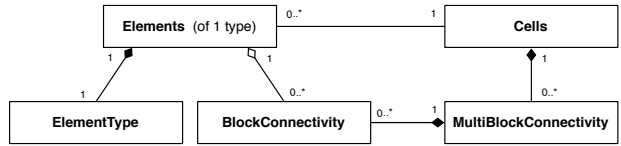

Figure 26: Mesh *Cells* diagram.

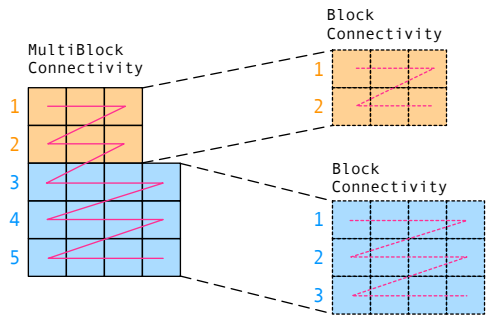

Figure 27: *BlockConnectivity* points to blocks of *MultiBlockConnectivity*. Zig-zag lines denote how the data is laid out contiguously in memory.

Although currently the mesh is composed of two-dimensional elements such as quadrilaterals and triangles, three-dimensional mesh elements such as hexahedra, tetrahedra, etc. are envisioned in the design and can be naturally embedded within the presented data structure. However, at least for the foreseeable future in NWP and climate applications, the vertical discretisation may be considered orthogonal to the horizontal discretisation due to the large anisotropy of physical scales in horizontal and vertical directions. Given a number of vertical levels, polygonal elements in the horizontal are then extruded to prismatic elements oriented in the vertical direction (e.g. [14]).

## 4.4 Parallelisation

Parallelisation in *Atlas* is achieved through distributing the *Mesh* into different partitions, each acting like a smaller mesh and each mesh partition $\mathcal{M}_\mathrm{p}$ is managed by one MPI task. The idea is to load-balance numerical computations and memory among the MPI tasks, meaning that every mesh partition has approximately the same number of elements, or the same number of nodes.

Looking back at the typical workflow on how to use *Atlas*, presented in Figure 2, we start with a *Grid* object that doesn't have any notion of parallelisation, and we want to end up with a distributed *Mesh* object. One approach could be to first generate the mesh from the grid on one MPI task with a *MeshGenerator* object (see Section 4.3), then call a partitioning algorithm on the mesh, and then distribute the mesh partitions to the other MPI tasks. This approach has major flaws in parallel efficiency as many MPI tasks are waiting for computations of the master MPI task to finish, and then wait to receive their mesh partition. Another approach would be to do a partitioning of the grid points before or during the mesh generation step, and only generate the mesh partitions using the grid points whose partitioning corresponds to the required MPI task. In principle this is applicable to both *UnstructuredGrid*s and *StructuredGrid*s. Currently *Atlas* has however only implemented such parallel enabled mesh generator for *StructuredGrid*s.

Examples of two meshes partitioned into different parallel regions using the *Equal-Regions* partitioning algorithm are illustrated in Figure 28.

Every mesh partition can be regarded as an independent mesh, but to allow for computational stencils that span from one mesh partition to the next, halo's that overlap are created between relevant mesh partitions. *Atlas* provides functionality to incrementally grow the overlap between mesh partitions by node-sharing elements. Figure 29 shows the overlap region generated for two such regions, as well as a so called "periodic overlap region" that can be used to treat the periodic East-West boundary as if it were an internal boundary between mesh partitions. Discrete field values present in overlap regions require synchronisation with values of neighbouring partitions for performing stencil operations. For this synchronisation, the mesh partition must be aware of how it fits inside the whole mesh. As shown in Figure 24, the *Nodes*, *Cells*, and *Edges* classes contain three fields, intended as discrete values, that provide exactly this awareness.

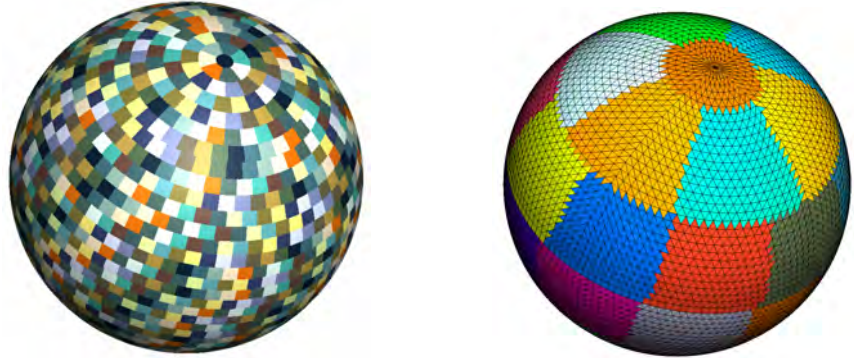

Figure 28: *EqualRegions* domain decomposition. Left: `O1280` mesh with ∼ 6.6 million nodes (∼ 9 km grid spacing) in 1600 partitions. Right: `O32` mesh with 5248 nodes (∼ 280 km grid spacing) in 32 partitions.

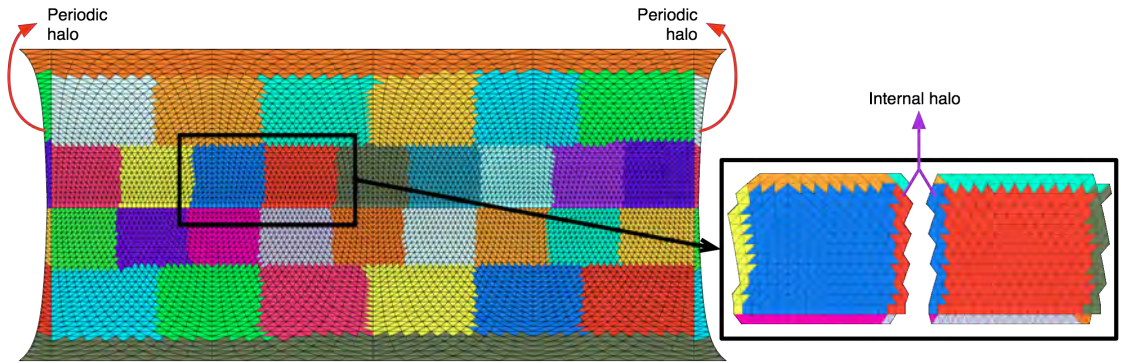

Figure 29: Parallel overlap regions or halo's shown for a `O32` mesh with 32 partitions.

- The field named *global_index* contains a unique global index or ID for each node or element in the mesh partition as if the mesh was not distributed. The global index is independent of the number of partitions.

- The field named *partition* contains the partition index that has ownership of the node or element. Nodes or elements whose partition does not match the partition index of the mesh partition are also called ghost nodes or ghost elements respectively. These ghost entities merely exist to facilitate stencil operations (such as derivatives) or to complete, for instance, a mesh element.

- The field named *remote_index* contains the location or local index of each node or element on the partition that owns it.

With the knowledge of *partition* and *remote_index*, it is possible to know, for each element or node, which partition owns it and at which local index therein. Usually the *Atlas*' user will not be aware of these three fields as they are required only for constructing *Atlas*' internal parallel communication capabilities.

Currently, *Atlas* provides two parallel communication classes that, given the three fields such as *partition*, *remote_index* and *global_index*, can apply parallel communication operations repeatedly as needed:

- The *GatherScatter* class implements the communication operation that gathers data from all MPI tasks to one MPI task, and vice versa: the communication operation that scatters or distributes all data from one MPI task to all MPI tasks.

- The *HaloExchange* class implements the communication operation that sends and receives data to and from MPI tasks containing nearest-neighbour partitions. This operation is typically required when synchronising halo's of ghost entities surrounding a domain partition.

These parallel communication classes form building blocks that provide parallel capabilities to the *FunctionSpace* class, which can manage the gathering, scattering or halo-exchanging of *Field*s.

## 4.5 FunctionSpace

The *FunctionSpace* class is introduced because a *Field* (Section 4.6) can be discretised on the computational domain in various ways: e.g. on a grid, on mesh-nodes, mesh-cell-centers or spectral coefficients. The representation of a given variable is intimately related to the spatial numerical discretisation strategy one wants to adopt (e.g. finite volume, spectral element, spectral transform, etc.). In addition to interpreting how a *Field* is discretised, the *FunctionSpace* also manages how the *Field* is parallelised and laid out in memory. It implements parallel operations such as gather and scatter, reduce-all, point-to-point communications, thus enabling the practical use of fields within parallel numerical algorithms.

In *Atlas*, the *FunctionSpace* concept, depicted in Figure 30, is implemented in a modular OO paradigm that allows adding as many different function spaces as required. This modularity allows third-party applications to extend the library with their own *FunctionSpaces* while still profiting from the parallelisation primitives provided by *Atlas* (highlighted in dashed blue). The currently implemented *FunctionSpace* classes include *NodeColumns*, *EdgeColumns*, *StructuredColumns* and *Spectral*:

- The *NodeColumns* function space class describes the discretisation of fields with values collocated at the nodes of the mesh, horizontally, and may have multiple layers defined in the vertical direction. Parallelisation is defined in the horizontal plane, so that complete vertical columns are available on each partition. The memory layout for fields defined using the *NodeColumns* function space is illustrated in Figure 31. A *HaloExchange* object and *GatherScatter* object are responsible for the necessary parallel operations (Section 4.4). The *NodeColumns* function space also implements some simple additional features, such as calculating global minimum and maximum values of fields as well as some global reduction computations such as arithmetic mean values.

- The *EdgeColumns* function space class describes the discretisation of fields with values collocated at the edges of the mesh, and may have multiple layers defined in a vertical direction. The various operations just described for the *NodeColumns* class are also available for this class.

- The *StructuredColumns* function space class describes the discretisation of distributed fields on a *Structured* grid object. Currently the *Structured* grid must be Gaussian (see Section 4.2) because the function space delegates its parallel primitives to a specific *Trans* object that only supports Gaussian grids. As the *Trans* object is an interface with an external library that implements spectral transformations, we do not report the details here, but it is a good example of interfacing with pre-existing high performance codes. In a future release the parallelisation will be generalised to use a *GatherScatter* object instead, which does not rely on having a Gaussian grid. A field described using this function space, like the two above, can also have vertical levels.

- The *Spectral* function space class describes a field in terms of vertical layers of horizontal spherical-harmonics (global spherical representation). The parallelisation (gathering and scattering) is again delegated to the *Trans* object.

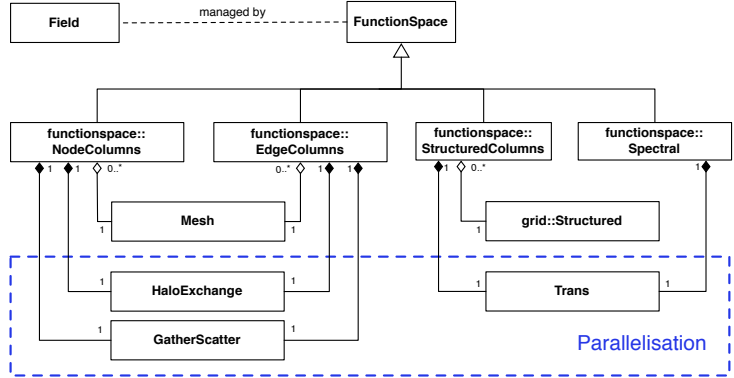

Figure 30: *FunctionSpace* implementations including building blocks required to interpret *Fields* and abstract parallelisation.

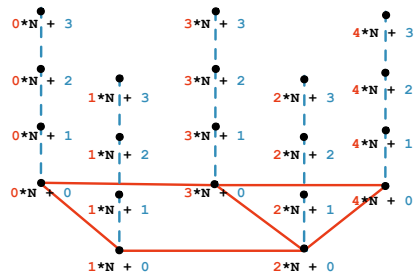

Figure 31: Memory layout for fields discretised using the *NodeColumns* function space. A vertical column is contiguous in memory, and can be indexed using direct addressing. *N* stands for the number of vertical layers.

Listing 8 and 9 are provided to help understand how a *FunctionSpace* can be used in practice to create a field, and perform a halo-exchange on this field. Listing 8 and 9 show both the C++ and the Fortran code, respectively.

```
NodeColumns functionspace( mesh, Halo(1) );
Field field = functionspace.createField<double>( field::levels(100) );
functionspace.haloExchange( field );
```

Listing 8: C++ *FunctionSpace* example use

```
type(atlas_functionspace_NodeColumns) :: functionspace
type(atlas_Field)                      :: field
functionspace = atlas_functionspace_NodeColumns( mesh, halo=1 )
field         = functionspace%create_field( atlas_real(8), levels=100 )
call functionspace%halo_exchange( field )
```

Listing 9: Fortran *FunctionSpace* example use

## 4.6 Field

The *Field* class contains the values of a full scalar, vector or tensor field. The *Field* values are stored contiguously in memory, and moreover they can be mapped to an arbitrary indexing mechanism to target a specific memory layout. The ability to adapt the memory layout to match for instance the most efficient data access patterns of a specific hardware is a key feature of *Atlas*. A *Field* also contains *Metadata* which stores simple information like a name, units, or other relevant information. The composition of the *Field* class is illustrated in Figure 32. A *Field* delegates the access and storage of the actual memory to an *Array* that accommodates memory storage on heterogeneous hardware[1]. If the *Field* is associated to a particular *FunctionSpace*, then the *Field* also contains a reference to it.

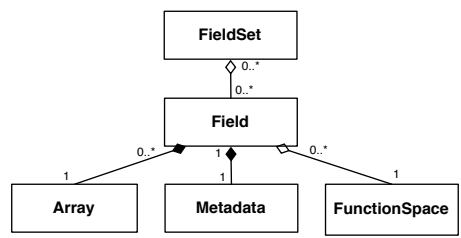

Figure 32: *Field* composition.

A *FunctionSpace*, as mentioned, permits the definition of parallel operations to be carried out on a given field. It defines a memory layout and is related to a particular spatial discretisation.

*Field*s can also be grouped together into one or more *FieldSet*s. They can then be accessed from the *FieldSet* by name or by index. In C++, access to the actual field data is via an `make_view<Value,Rank>()` construct that creates a *view* of the field data with a multi-dimensional indexing accessor. In Fortran, the data is directly accessed through the multi-dimensional array intrinsics of the language. Practical use of the *Field*, both using C++ and Fortran, is given Listings 10 and 11.

---

[1]The *Array* is responsible to synchronise data across the device (e.g. a GPU) and the host (e.g. a CPU).

```cpp
FieldSet fields;
fields.add( functionspace.createField<double>(field::name("temperature"),
    field::levels(nb_levels)) );
fields.add( functionspace.createField<double>(field::name("pressure"),
    field::levels(nb_levels)) );

Field field_T = fields["temperature"];
Field field_P = fields["pressure"];

// Create (2D) views of the fields to access the data
auto T = make_view<double,2>(field_T);
auto P = make_view<double,2>(field_P);
for( size_t jnode=0; jnode<nb_nodes; ++jnode ) {
    for( size_t jlev=0; jlev<nb_levels; ++jlev) {
    // T(jnode,jlev) = ...
    // P(jnode,jlev) = ...
    }
}
```

Listing 10: C++ *Field* base class

More detail on the *Array* and *ArrayView* class can be found in Section 5.

```fortran
type(atlas_FieldSet) :: fields
type(atlas_Field)    :: field_T, field_P
real(8), pointer     :: T(:,:), P(:,:)
fields = atlas_FieldSet()
call fields%add( functionspace%create_field(kind=atlas_real(8),name="
    temperature",levels=nb_levels) )
call fields%add( functionspace%create_field(kind=atlas_real(8),name="
    pressure",levels=nb_levels) )

field_T = fields%get("temperature")
field_P = fields%get("pressure")

call field_T%data(T)
call field_P%data(P)

do jnode=1,nb_nodes
    do jlev=1,nb_levels
    ! T(jlev,jnode) = ...
    ! P(jlev,jnode) = ...
    enddo
enddo
```

Listing 11: Fortran *Field* base class

## 4.7   Mathematical Operations

Many NWP and climate models contain algorithms to perform a variety of mathematical operations on fields such as computing derivatives or integrals. These operations are common to various applications, and relate closely to certain spatial discretisations or function spaces. *Atlas* provides implementations for some of these operations given a field that is compatible with the related *FunctionSpace* (Section 4.5). Figure 33 sketches the philosophy adopted by *Atlas* regarding how to provide these operators. The concrete implementation of the *Method* concept uses the *FunctionSpace* and *Field* classes, both required to generate a concrete numerical method. *Atlas* currently provides a *fvm::Method* class, which contains everything required to construct mathematical operators using an edge-based finite volume scheme [15]. A concrete *fvm::Nabla* operator then implements the actual numerical algorithm using the *fvm::Method*. Listing 12 details the practical construction of the *fvm::Method* and how the gradient of a scalar field defined in *NodeColumns* is constructed. Note that this implementation can compute gradients of three-dimensional fields (with vertical levels), but only computes the horizontal components.

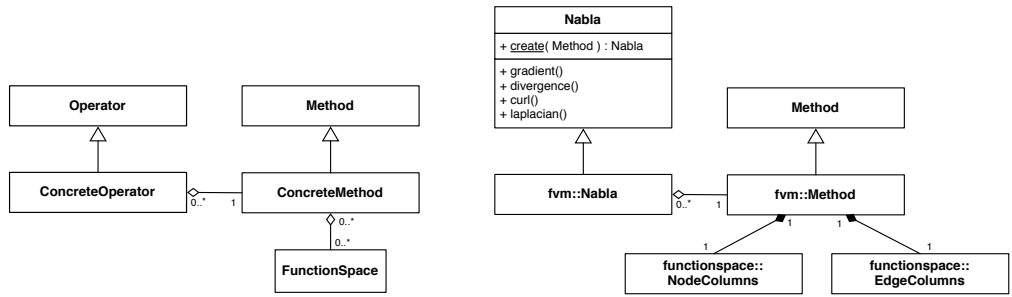

Figure 33: Left: general design of numerical operators. Right: Derivative, divergence, curl, and Laplacian implemented in the *Nabla* vector operator specific for a finite volume *Method* [15].

```cpp
fvm::Method method( mesh );
Nabla nabla( method );

Field scalar_field   = method.nodeColumns().createField<double>(
    field::levels(100) );
Field gradient_field = method.nodeColumns().createField<double>(
    field::levels(100), field::variables(2) );

/* ... code missing that sets up the scalar_field ... */

nabla.gradient(scalar_field,gradient_field);
```

Listing 12: C++ numerical operator Nabla that computes the gradient of a scalar field

# 5   Accelerator Support

Atlas is used as the abstraction layer of the underlying grid used for implementing numerical operators of many of the dwarfs proposed in the ESCAPE project. In order to support the port of the different dwarfs to heterogeneous architectures the Atlas library needs to be extended. In particular for NVIDIA GPUs, the memory space of the accelerator is separated from the CPU that manages the memory. In this section we describe the developments performed in order to support GPUs by encapsulating the data-management of the Atlas mesh and field data structures into the application-programming-interface(API) of Atlas.

In a first phase, different approaches and strategies to support Atlas data structures for accelerators were studied. Since the ESCAPE DSL (based on the GridTools library) will also use the Atlas mesh data structures to implement numerical operators on irregular grids on the sphere (Task 2.3), the interoperability of the Atlas data structures and ESCAPE DSL library is an important aspect.

Three possible strategies were evaluated:

1. Allocate mirror storages for the accelerator memory within the existing data structures using the CUDA API for memory management and provide functions to synchronise the CPU and accelerator memory spaces. In order to inter-operate with the DSL a converter between GridTools and Atlas is required.

2. Replace the existing Atlas data structures by GPU capable GridTools storages.

3. Hybrid solution where both options, the Atlas native data structures and GridTools storages are supported.

Using the GridTools storage has the advantage that its storage management framework already solves the problem of supporting storages in the accelerator memory and provides an API to operate on them and synchronise the CPU and accelerator copies. Additionally the framework allows to flexibly choose the most efficient memory layout for different computing architectures. Finally an integration of GridTools storage as the underlying data management layer for Atlas data structures will provide a high level of interoperability with the DSL, as it uses the same storages. Therefore option 3 was chosen, since it benefits from the aforementioned advantages and at the same time retains the native Atlas storage implementation and does not enforce a dependency on the GridTools library.

## 5.1 GridTools storage layer

The GridTools storage module provides flexible data structures for storing fields on a grid with support for GPU accelerators.

Usually storages of programming languages like C++ or Fortran do not allow to specify the memory layout of the space (and extra) dimensions of a field. This is a crucial functionality for performance portability, since different algorithm motifs and computing architectures require different memory layout for an efficient access to the memory and to increase the data locality aspects of the algorithm.

The C++ data structures of GridTools are very general and allow to customise properties like dimensionality of the field, memory layout, alignment, accelerator support, etc. Listing 13 shows the example of a creation of a customised GridTools storage. The memory layout is abstracted in this case by the CUDA backend which chooses the optimal layout for GPUs.

```cpp
using storage_info_t = storage_traits< Cuda >::storage_info_t<
    3,                  // Rank
    halo<2,2,0>         // Halo of size 2 for indices i,j
  >;
using data_store_t = storage_traits< Cuda >::data_store_t<
    double,             // Data type
    storage_info_t      // Data storage info
  >;

// Horizontal wind field with Ni = Nj = 128 with 2 components (u,v)
data_store_t wind(128,128,2);
```

Listing 13: C++ Example of Gridtools storage API. It shows the creation of a storage for a rank-3 array of double precision and a halo of 2 grid points in the `i` and `j` dimensions.

This low-level GridTools storage framework is managed by the Atlas data structures, whose API abstracts the underlying implementation (native or GridTools storages) and allows to access and synchronise the GPU and CPU memory spaces. Further details will be provided in the following sections.

For more information on the low-level GridTools storage capabilities, refer to the GridTools developments [6].

## 5.2 Atlas and GridTools storage integration

The two main Atlas data structures that need to be supported for accelerators are the *Array*, used by the Atlas *Field* (Section 4.6) and the connectivity classes used by the *Mesh* (Section 4.3). Both type of Atlas data structures have now GPU support by means of the GridTools storage framework and provide an API to manage the GPU memory space of the fields and connectivity tables.

Additionally the `make_view` constructs of Atlas support now the creation of specific array views for the CPU and the GPU device. The Atlas *ArrayView* is used to interpret a given storage as N-dimensional array by providing a parenthesis operator for accessing the data, similar to the Fortran syntax to access N-dimensional arrays. The GridTools storage has the advantage that the memory layout can be customised to provide an optimal layout for a specific computing architectures.

The UML class diagram for the *Array* class and its relation to *ArrayView* is shown in Figure 34. It is shown how the *Array* abstracts its implementation to use either the *Atlas* native data storage or the GridTools storage.

---

**Note**

To take advantage of the GPU related capabilities, the *GridToolsDataStore* implementation needs to be selected with the CUDA GPU backend. This is achieved via compile time options `-DENABLE_GRIDTOOLS_STORAGE=ON -DENABLE_GPU=ON` (see Section 3.3).

---

Listing 14 shows and example of creation and use of a CPU array view. Similarly Listing 15 demonstrate the creation of a storage that is cloned to the GPU and the creation and use, within a CUDA kernel, of a device view.

```
Array* ds = Array::create<double>(nb_nodes, nb_levels);
// Create a host view to interpret the Array as a 2D storage of doubles
auto hv = make_host_view<double, 2>(*ds);
for ( size_t jnode = 0; jnode < nb_nodes; ++ jnode ) {
  for ( size_t jlev = 0; jlev < nb_levels; ++ jlev ) {
    // hv(jnode, jlev) = ...
  }
}
```

Listing 14: C++ Example of creation of a host array view

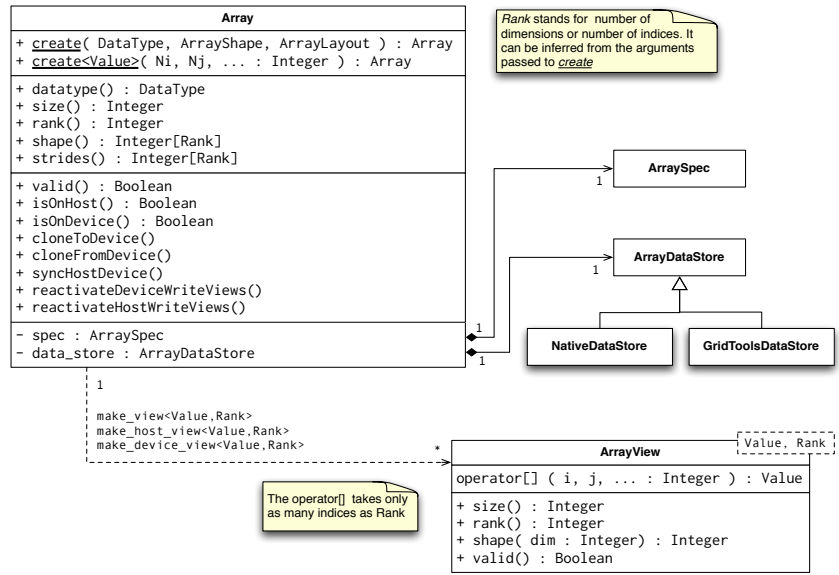

Figure 34: UML diagram for the *Array* class and its relation to the *ArrayView* class

```cpp
__global__
void kernel_ex(ArrayView<double, 2> dv, size_t nb_levels)
{
  for(size_t jlev=0; jlev < nb_levels; ++jlev)
    dv(threadIdx.x, 3) = ...;
}

// Create an Atlas array
Array* ds = Array::create<double>(nb_nodes, nb_levels);

// Synchronise the GPU device copy of the array
ds->cloneToDevice();

// Create a (2D) view that can be used from a GPU kernel
auto dv = make_device_view<double, 2>(*ds);

// GPU kernel computation that uses the array view
kernel<<<functionspace.nb_nodes(),1 >>>(dv, nb_levels);
cudaDeviceSynchronize();

// Synchronise the CPU copy of the array
ds->cloneFromDevice();

// create a host view to interpret the Array as a 2D storage of doubles
auto hv = make_host_view<double, 2>(*ds);

for ( size_t jnode = 0; jnode < nb_nodes; ++ jnode ) {
  for ( size_t jlev = 0; jlev < nb_levels; ++ jlev ) {
    // check the values computed
    // if( hv(jnode, jlev) == ... ) ...
  }
}
```

Listing 15: C++ Example of creation of a device array view and use of a GPU kernel using CUDA

**Synchronisation protections**

Once Atlas can create views of an array in multiple memory spaces (host and GPU device), the computation can lead to invalid states of the array, if both the host CPU the GPU views update their corresponding memory space without synchronising them accordingly, as shown in the following example

```cpp
// Create an Atlas array
Array* ds = Array::create<double>(nb_nodes, nb_levels);

// Synchronise the GPU device copy of the array
ds->cloneToDevice();

// Create a view that can be used from the CPU
auto hv = make_host_view<double, 2>(*ds);

// Create a view that can be used from a GPU kernel
auto dv = make_device_view<double, 2>(*ds);

// Modify the host view
for ( size_t jnode = 0; jnode < nb_nodes; ++ jnode ) {
  for ( size_t jlev = 0; jlev < nb_levels; ++ jlev ) {
    hv(jnode, jlev) = 0;
  }
}

// Modify the device view using a CUDA kernel
kernel<<<nb_nodes,1 >>>(dv, nb_levels);
cudaDeviceSynchronize();

// At this point the two memory spaces are in a different state,
// invalidating both views
// --> dv.valid() == false
// --> dh.valid() == false
```

Listing 16: C++ Example of creation of two views (host and device) that modify concurrently their memory spaces leading to an inconsistent stage label

The current state of a view can be checked anytime with the *valid* method.

In order to support multiple views that can coexist in the same scope avoiding invalid states, Atlas gives the possibility to create read only views, that will never invalidate the state of other existing views, since they do not allow to modify their data. Listing 17 shows an example of a valid use and coexistence of multiple views by making use of read only views.

```
// Create an Atlas array
Array* ds = Array::create<double>(nb_nodes, nb_levels);

// Create a read only host view
auto hv = make_host_view<double, 2, true>(*ds);

// Create a write device view
auto dv = make_device_view<double, 2, false>(*ds);
```

Listing 17: C++ Example of creation and coexistence of multiple views by creating Atlas read only views (last optional template parameter of the make_view constructs)

## 5.3 Fortran fields on accelerators

One of the main functionalities of Atlas is the support of Fortran bindings so that Fortran numerical operators can be implemented using the underlying Atlas data structures for meshes and fields.

Therefore the GPU capable storages and API to manage the data has been forwarded to the Fortran API of Atlas. In order to port Fortran numerical operators to the GPU, one of the programming models employed by the ESCAPE project is OpenACC.

OpenACC is a directive based approach to port Fortran operators to accelerators that allows to retain the original implementation by using directives (comments in the Fortran code) to instruct the OpenACC which loops should be parallelized in the GPU.

In order to support the OpenACC development, Atlas connects the GPU allocated pointers of the fields with the OpenACC gpu pointers. Listing 18 shows an example of how to use the GPU storages of Atlas to implement an OpenACC kernel on the GPU.

```fortran
type(atlas_Field) :: field1
type(atlas_Field) :: field2
real(8), pointer :: v1(:,:)
real(8), pointer :: v2(:,:)

field1 = atlas_Field(kind=atlas_real(8),shape=[n,n])
field2 = atlas_Field(kind=atlas_real(8),shape=[n,n])

call field1%clone_to_device()
call field2%clone_to_device()

call field%device_data(v1)
call field%host_data(v2)

!acc data present(v1) copyin(v2)
!$acc kernels
do j=1,n
  do i=1,n
    v1(i,j) = v2(i,j)+ 42.
  enddo
enddo
!$acc end kernels
!$acc end data
```

Listing 18: Fortran Example of an OpenACC kernel operating on Atlas fields

# 6 Conclusions

The *Atlas* C++/ Fortran library provides flexible data structures for both structured and unstructured meshes and is intended to be applied in NWP or Climate modelling codes.

One of the key data structure components used in a model is the *Field* that holds the actual data of a field variable. During the course of the ESCAPE project, the *Field* concept has been extended to be GPU aware so that new algorithmic approaches previously not possible can now be coded based on *Atlas*. One approach is based on the GridTools Domain Specific Language, which is going to be applied to several ESCAPE dwarfs. Thanks to the ESCAPE project, the use of *Atlas* fields is now capable of being used easily in a "host-device" combination of heterogeneous hardware. Including the expertise of partners using GPU's and developing GPU hardware at this early development stage of *Atlas* has been vital in developing a future-proof data structure library.

The grid and mesh generation facilities in *Atlas* have been extended as part of Deliverable D4.4 to include regional grids defined in projected coordinates ($x$,$y$), rather than geospherical coordinates (longitude,latitude). Thanks to the ESCAPE project, the Limited Area Modelling (LAM) community was involved at the early development stage of *Atlas*. It would have proved a much more difficult task to redesign *Atlas* after the library would have matured without the involvement of the LAM ESCAPE partners.

A new stable *Atlas* release for ESCAPE's further dwarf developments is now established with this deliverable, and has been tested and compiled with various compilers and computer architectures.

# 7 References

[1] *ESCAPE Deliverable D1.1*, `http://www.hpc-escape.eu/media-hub/escape-pub/escape-deliverables`, Accessed: 31-03-2017.

[2] *ESCAPE Deliverable D1.2: Batch 2: Definition of novel Weather & Climate Dwarfs, provision of prototype implementations and dissemination to other WPs*, `http://www.hpc-escape.eu/media-hub/escape-pub/escape-deliverables`, Available: 31-12-2017.

[3] *ESCAPE Deliverable D4.4: Atlas extension for LAM use*, `http://www.hpc-escape.eu/media-hub/escape-pub/escape-deliverables`, Accessed: 31-03-2017.

[4] *ESCAPE Deliverable D2.4: Domain-specific language (DSL) for dynamical cores on unstructured meshes/structured grids*, `http://www.hpc-escape.eu/media-hub/escape-pub/escape-deliverables`, Available: 31-12-2017.

[5] *NVIDIA CUDA toolkit documentation*, `http://docs.nvidia.com/cuda/`, Accessed: 31-03-2017.

[6] T. Gysi, C. Osuna, O. Fuhrer, M. Bianco, and T. C. Schulthess, "Stella: A domain-specific tool for structured grid methods in weather and climate models", in *Proceedings of the International Conference for High Performance Computing, Networking, Storage and Analysis*, ser. SC '15, Austin, Texas: ACM, 2015, 41:1–41:12, ISBN: 978-1-4503-3723-6. DOI: `10.1145/2807591.2807627`. [Online]. Available: `http://doi.acm.org/10.1145/2807591.2807627`.

[7] *ESCAPE Deliverable D5.6: Establish software collaboration platform*, `http://www.hpc-escape.eu/media-hub/escape-pub/escape-deliverables`, Accessed: 31-03-2017.

[8] *Unified Modeling Language main webpage*, `http://www.uml.org`, Accessed: 31-03-2017.

[9] M. Hortal and A. Simmons, "Use of reduced Gaussian grids in spectral models.", *Monthly Weather Review*, vol. 119, pp. 1057–1074, 1991.

[10] S. Malardel, N. Wedi, W. Deconinck, M. Diamantakis, C. Kühnlein, G. Mozdzynski, M. Hamrud, and P. Smolarkiewicz, "A new grid for the IFS", *ECMWF Newsletter*, vol. 146, pp. 23–28, 2016.

[11] W. Deconinck, M. Hamrud, C. Kühnlein, G. Mozdzynski, P. Smolarkiewicz, J. Szmelter, and N. Wedi, "Accelerating extreme-scale numerical weather prediction", in *Parallel Processing and Applied Mathematics*, Springer, 2016, pp. 583–593.

[12] P. Leopardi, "A partition of the unit sphere into regions of equal area and small diameter", *Electronic Transactions on Numerical Analysis*, vol. 25, no. 12, pp. 309–327, 2006.

[13] G. Mozdzynski, "A new partitioning approach for ECMWF's integrated forecasting system (IFS)", in *Proceedings of the Twelfth ECMWF Workshop: Use of High Performance Computing in Meteorology*, vol. 273, World Scientific, 2007, pp. 148–166.

[14] A. MacDonald, J. Middlecoff, T. Henderson, and L. J.-L., "A general method for modeling on irregular grids", *High Performance Computing Applications*, vol. 25, no. 1, pp. 392–403, 2011.

[15] P. K. Smolarkiewicz, W. Deconinck, M. Hamrud, C. Kühnlein, G. Mozdzynski, J. Szmelter, and N. P. Wedi, "A finite-volume module for simulating global all-scale atmospheric flows", *Journal of Computational Physics*, vol. 314, pp. 287–304, 2016.

# Document History

| Version | Author(s) | Date | Changes |
|---------|-----------|------|---------|
| 0.1 | W. Deconinck | 9/01/17 | Initial layout |
| 0.2 | W. Deconinck | 24/02/17 | Design concepts |
| 0.3 | W. Deconinck | 05/03/17 | Installation instructions |
| 0.4 | W. Deconinck | 19/03/17 | LAM contributions |
| 0.5 | C. Osuna | 20/03/17 | Accelerator support |
| 0.6 | W. Deconinck | 21/03/17 | Version for internal review |
| 1.0 | W. Deconinck | 30/03/17 | Final version |

# Internal Review History

| Internal Reviewers | Date | Comments |
|--------------------|------|----------|
| Daan Degrauwe, Joris Van Bever | 25/03/17 | Approved with comments |
| Carlos Osuna | 27/03/17 | Approved with comments |

# Effort Contributions per Partner

| Partner | Efforts |
|---------|---------|
| ECMWF | 6 pm |
| RMI | 0.5 pm |
| MeteoSwiss | 5.5m |
| Total | 12m |

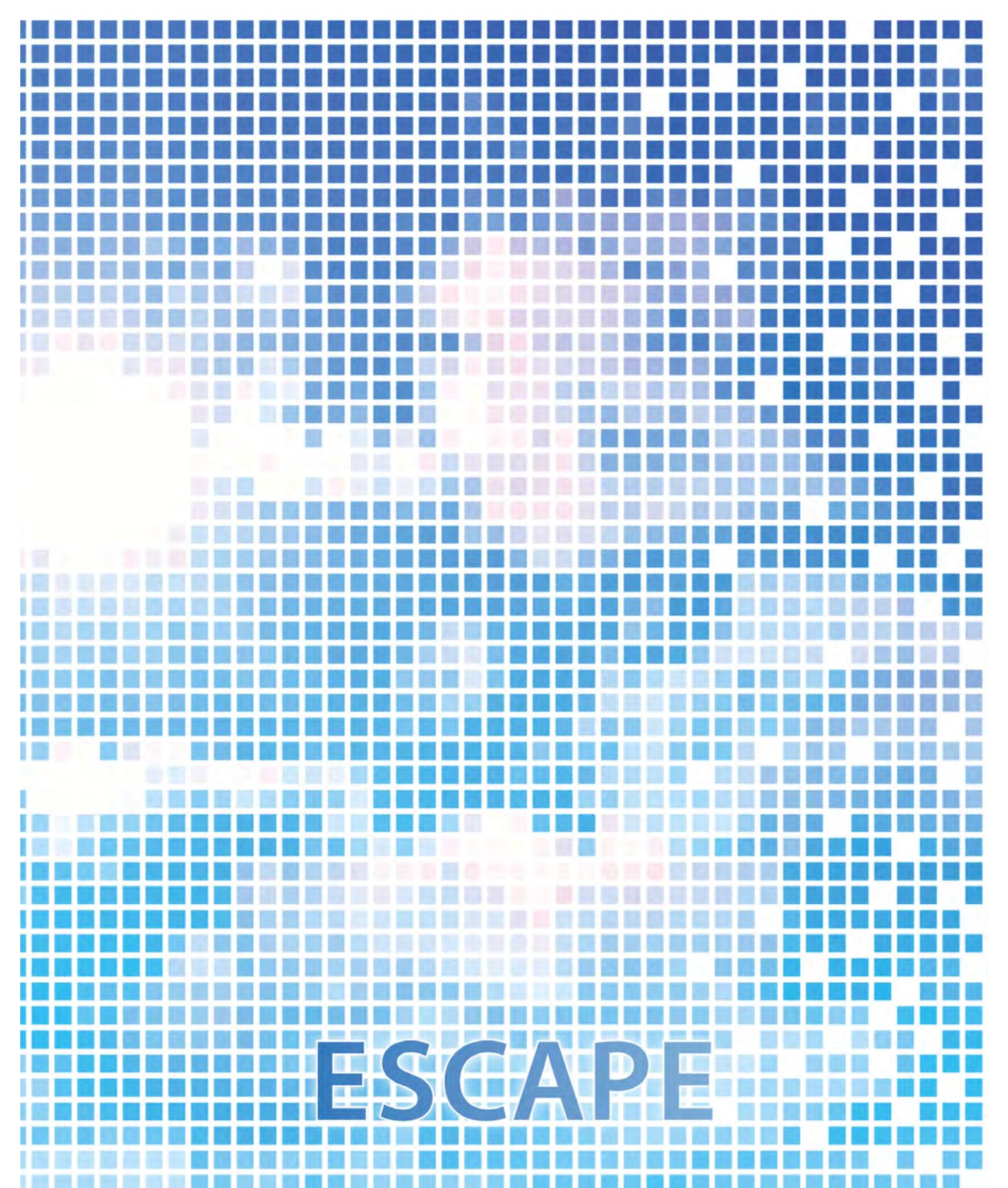